\documentclass[10pt,journal,compsoc]{IEEEtran}

\usepackage{ifpdf}        %
\usepackage{amsmath}      %
\usepackage{algorithmic}  %
\usepackage{array}        %
\usepackage{url}          %

\ifpdf
\else
\fi

\ifCLASSOPTIONcompsoc
  \usepackage[nocompress]{cite}
\else
  \usepackage{cite}
\fi

\ifCLASSINFOpdf
   \usepackage[pdftex]{graphicx}
   \graphicspath{{figures/}{pictures/}{images/}{./}}  %
   \DeclareGraphicsExtensions{.pdf,.png,.jpg,.jpeg}   %
\else
   \usepackage[dvips]{graphicx}
   \graphicspath{{figures/}{pictures/}{images/}{./}}  %
   \DeclareGraphicsExtensions{.eps}                   %
\fi

\ifCLASSOPTIONcompsoc
 \usepackage[caption=false,font=footnotesize,labelfont=sf,textfont=sf]{subfig}
\else
 \usepackage[caption=false,font=footnotesize]{subfig}
\fi

\ifCLASSOPTIONcaptionsoff
 \usepackage[nomarkers]{endfloat}
\let\MYoriglatexcaption\caption
\renewcommand{\caption}[2][\relax]{\MYoriglatexcaption[#2]{#2}}
\fi

\usepackage{tabu}                %
\usepackage{booktabs}            %
\usepackage{enumitem}            %
\usepackage[dvipsnames]{xcolor}  %
\usepackage[normalem]{ulem}      %
\usepackage{hyperref}            %

\definecolor{mred}{rgb}{.80,.12,.30}
\definecolor{tableau1}{RGB}{31, 119, 180}
\definecolor{tableau2}{RGB}{255, 127, 14}
\definecolor{tableau3}{RGB}{44, 160, 44}
\definecolor{tableau10}{RGB}{23, 190, 207}

\newif\ifauthornotes
\newif\ifstrike
\newif\ifadd
\newif\ifreplace
\newif\iftodo

\newcommand{\strike}[1]{\ifstrike{\color{mred}{\texorpdfstring{\sout{#1}}{#1}}}\fi}

\newcommand{\add}[1]{\ifadd{\leavevmode\color{magenta}{#1}}\else{#1}\fi}

\begin{document}

\title{KnowledgeVIS: Interpreting Language Models by Comparing Fill-in-the-Blank Prompts}

\author{%
  Adam~Coscia and Alex~Endert%
  \IEEEcompsocitemizethanks{%
    \IEEEcompsocthanksitem Adam Coscia and Alex Endert are with Georgia Institute of Technology. Emails: \{acoscia6, endert\}@gatech.edu.
  }%
  \thanks{\textcolor{red}{This manuscript is a Preprint - Accepted, to Appear: IEEE Transactions on Visualization and Computer Graphics, DOI: 10.1109/TVCG.2023.3346713}}
}%

\IEEEtitleabstractindextext{%
  \begin{abstract}
    Recent growth in the popularity of large language models has led to their increased usage for summarizing,\strike{ translating,} predicting, and generating text, making it vital to help researchers and engineers understand how and why they work.
    We present \textit{KnowledgeVIS}, a human-in-the-loop visual analytics system for interpreting language models using fill-in-the-blank sentences as prompts.
    By comparing predictions between sentences, \textit{KnowledgeVIS} reveals learned associations that intuitively connect what language models learn during training to natural language tasks downstream\strike{. Tightly integrated views in KnowledgeVIS help}\add{, helping} users create and test multiple prompt variations, analyze predicted words using a novel semantic clustering technique, and discover insights using\strike{ expressive and} interactive visualizations.
    Collectively, these visualizations help users identify the likelihood and uniqueness of individual predictions, compare sets of predictions between prompts, and summarize patterns and relationships between predictions across all prompts.
    We demonstrate the capabilities of \textit{KnowledgeVIS} with feedback from six NLP experts as well as three different use cases: (1) probing biomedical knowledge in two domain-adapted models; and (2) evaluating harmful identity stereotypes and (3) discovering facts and relationships between three general-purpose models.\strike{KnowledgeVIS shows how human-in-the-loop visual analytics can help researchers interpret language models by providing insight into what they have learned.}
  \end{abstract}

}

\maketitle

\IEEEdisplaynontitleabstractindextext

\IEEEraisesectionheading{
  \section{Introduction}
  \label{sec:introduction}
}

\IEEEPARstart{L}{}arge language models (LLMs) such as BERT \cite{Devlin:2019:BERT} and GPT-3 \cite{Brown:2020:GPT3} have seen significant improvements in performance on natural language tasks \cite{Srivastava:2023:BeyondImitationGame}, enabling them to help people answer questions, generate essays, summarize long articles, and more.
Yet understanding what these models have learned and why they work is still an open challenge \cite{Rogers:2020:Bertology, Srivastava:2023:BeyondImitationGame}\strike{, one that is critical for enabling natural language processing (NLP) researchers to improve model development; e.g., by mitigating harmful stereotypes \cite{Abid:2021:AntiMuslimGPT}}.
In particular, how learned text representations in BERT-based language models generalize to natural language tasks remains unclear \cite{Roberts:2020:KnowledgeLMParameters}.
\add{For natural language processing (NLP) researchers and engineers who increasingly train and deploy LLMs as ``black boxes'' for generating text, exploring how learned behaviors during training manifest in downstream tasks can help them improve model development; e.g., by surfacing harmful stereotypes \cite{Abid:2021:AntiMuslimGPT}.}
To help researchers and engineers close the gap between what BERT-based models have learned and how they perform on downstream tasks, we can use prompts formatted as natural language tasks \cite{Liu:2023:PromptingSurvey}.
For example, Table~\ref{tab:prediction_data} shows BERT's predictions for multiple fill-in-the-blank sentences that test for conceptual reasoning capabilities, connecting model performance to learned text representations.

Our aim is to utilize fill-in-the-blank sentences for interpreting BERT-based language models by revealing learned associations.
Much of the existing work seeks to automatically extract, augment, and test individual template sentences against a manually curated ``gold standard'' \cite{Jiang:2020:LPAQA, Shin:2020:Autoprompt, Qin:2021:LearningHowToAsk, Zhong:2021:Optiprompt}.
However, these quantitative benchmarks miss an opportunity for injecting a researcher's intuition and domain expertise into evaluating model performance \cite{Bender:2021:StochasticParrots}.
A human-in-the-loop solution can foster human and LLM interaction by continuously integrating feedback into the process of model development \cite{Wang:2021:HITLNLP}.
Further, testing one prompt at a time can limit the interpretability of LLMs by failing to surface associations or producing contradictory predictions \cite{Rogers:2020:Bertology, Warstadt:2019:AnalysisBERTKnowledge, Elazar:2021:ParaRel}.
Guiding the user through exploring multiple prompts at once can reveal insights and patterns that further improve our understanding of BERT-based models.
To realize these goals, a successful solution should help users quickly format and test multiple prompt variations simultaneously, structure sets of predicted words to make them easier to parse, and present the data at several levels of detail.

\begin{table}[!t]
  \setlength{\tabcolsep}{4.9pt}
  \renewcommand{\arraystretch}{1.0}
  \caption{Example Data Set Visualized by \textit{KnowledgeVIS}}%
  \label{tab:prediction_data}
  \scriptsize
  \centering
  \begin{tabu}{r *{3}{l}}
    \toprule
      prompt & prediction & probability & cluster\footnotemark \\ 
    \midrule 
      \begin{tabu}{*{2}{r}} You are likely to find a \textbf{snake} \\ in a \_\_\_ \end{tabu} & field & 0.066 & \color{tableau3}{physical entity} \\
      \begin{tabu}{*{2}{r}} One effect of \textbf{exercising} \\ is feeling \_\_\_ \end{tabu} & better & 0.296 & \color{tableau2}{abstraction} \\
      \begin{tabu}{*{2}{r}} You could be \textbf{sick} because \\ you are \_\_\_ \end{tabu} & pregnant & 0.209 & \color{tableau10}{condition} \\
      \begin{tabu}{*{2}{r}} If you want to \textbf{learn} then \\ you need a \_\_\_ \end{tabu} & teacher & 0.122 & \color{tableau3}{physical entity} \\
    \bottomrule
  \end{tabu}
  \vspace{-1em}
\end{table}

\footnotetext{We generate clusters based on shared taxonomic and semantic similarity of tokens, described in Sect.~\ref{sec:clustering_predictions}.}

We present \textit{KnowledgeVIS}, a human-in-the-loop visual analytics system for comparing fill-in-the-blank prompts to uncover associations from learned text representations.
\textit{KnowledgeVIS} helps users create effective sets of prompts, probe multiple types of relationships between words, test for different associations that have been learned, and find insights across several sets of predictions for any BERT-based language model.
It does so through a tight integration of multiple coordinated views.
First, we designed an intuitive visual interface that structures the query process to encourage both creativity and rapid prompt generation and testing.
Then, to reduce the complexity of the prompt prediction space, we developed a novel clustering technique that groups predictions by semantic similarity.
Finally, we provided several expressive and interactive text visualizations to promote exploration and discovery of insights at multiple levels of data abstraction: a heat map; a set view inspired by parallel tag clouds \cite{Collins:2009:ParallelTagClouds}; and scatterplot with dust-and-magnet positioning of axes \cite{Soo:2005:DustAndMagnet}.
Collectively, these visualizations help the user identify the likelihood and uniqueness of individual predictions, compare sets of predictions between prompts, and summarize patterns and relationships between predictions across all prompts.

To validate our approach, we demonstrate the capabilities of \textit{KnowledgeVIS} with three use cases and an expert evaluation.
First, we reveal learned biomedical associations in two domain-adapted LLMs, SciBERT \cite{Beltagy:2019:SciBERT} and PubMedBERT \cite{Gu:2021:PubMedBERT}.
Second, we uncover harmful learned identity stereotypes between two general-purpose LLMs, BERT \cite{Devlin:2019:BERT} and RoBERTa \cite{Liu:2019:RoBERTa}.
Third, we elicit and evaluate different world knowledge (i.e. commonsense relationships) learned by a large and a small general-purpose LLM, BERT \cite{Devlin:2019:BERT} and DistilBERT \cite{Sanh:2019:DistilBERT} respectively.
Finally, we conduct an evaluation with six academic NLP researchers and engineers.
\textit{KnowledgeVIS} surfaced several new insights for the domain experts, including how BERT-based language models handle parts of speech, transitivity, and semantic roles, as well as learned cultural and religious associations that biased predictions in unexpected ways.
Our participants wanted to evaluate their own models using \textit{KnowledgeVIS} and would recommend our interface to anyone interested in ``opening the black box of how [LLMs] work''.

In summary, our paper contributes: (1) an open-source\footnote{\url{https://github.com/AdamCoscia/KnowledgeVIS}} visual analytics system, \textit{KnowledgeVIS}, that implements text visualization techniques for comparing fill-in-the-blank prompts that reveal associations from learned text representations in BERT-based language models; (2) a novel taxonomy-based technique for semantically clustering prompt predictions; and (3) three use cases and an expert evaluation showing how \textit{KnowledgeVIS} helps NLP researchers interpret BERT-based language models.

\section{Related Work}
\label{sec:related_work}

\subsection{Modeling Language with Transformers}
\label{sec:language_modeling}
In this paper, we focus on BERT-based language models because their transformer architecture and masked language modeling pre-training easily adapt to our fill-in-the-blank prompts.
Language models learn to model the probability of a token occurring in a sequence, e.g., a word in a sentence.
Transformers extend this by encoding and decoding all input words in a sentence, generating weights that map \textit{attention} from an output to the most relevant inputs \cite{Vaswani:2017:AttentionIsAllYouNeed}.
These attention weights are attributed with storing factual and linguistic knowledge needed to perform natural language tasks \cite{Roberts:2020:KnowledgeLMParameters}.
With this architecture, transformer-based language models can pre-train via self-supervised learning using large-scale unlabeled document corpora on a variety of tasks.
BERT-based language models pre-train on masked language modeling, or repeatedly removing and predicting (masking) tokens in a finite sequence \cite{Devlin:2019:BERT}.
Thus, we can elicit learned text representations in the attention weights of masked language models simply by mirroring their pre-training task using fill-in-the-blank sentences as prompts \cite{Liu:2023:PromptingSurvey}.
We test the capabilities of our fill-in-the-blank prompt approach with BERT base and four other pre-trained BERT-based models --- RoBERTa \cite{Liu:2019:RoBERTa}, DistilBERT \cite{Sanh:2019:DistilBERT}, SciBERT \cite{Beltagy:2019:SciBERT}, and PubMedBERT \cite{Gu:2021:PubMedBERT}.

\subsection{Probing Language Models With Prompts}
\label{sec:prompt_based_learning}
Transformers are typically pre-trained on various tasks (Sect.~\ref{sec:language_modeling}) and then fine-tuned using task-specific objective functions.
Prompting instead emulates pre-training by formatting the downstream objective as a natural language task \cite{Liu:2023:PromptingSurvey}.
Consider the example prompts in Table~\ref{tab:prediction_data}.
Instead of creating an objective function and labeled training data for a large language model (LLM) to predict conceptual relationships, we reformulate the task as one that BERT-based language models have seen already --- a fill-in-the-blank sentence as a prompt.

Prior work in probing LLMs using prompts involves eliciting and interpreting learned syntactic, semantic, and world knowledge \cite{Rogers:2020:Bertology}.
For example, Petroni et al. \cite{Petroni:2019:LAMA} showed that LLMs can be queried using fill-in-the-blank sentences as prompts (e.g., asking BERT to complete the sentence, ``The capital of France is \_'' results in BERT responding with ``Paris'') to elicit relational knowledge (i.e. the relationship ``capital of'' between France and Paris).
They posit knowledge probing estimates a lower-bound on knowledge contained in a model's internal representations.
Researchers have developed both manual and automated approaches to finding more effective prompts to raise the lower bound.
Jiang et al. created LPAQA \cite{Jiang:2020:LPAQA} using both mining-based and paraphrasing-based methods to generate new prompts for existing relations.
Shin et al. \cite{Shin:2020:Autoprompt} created a model, AUTOPROMPT, that searches for specific discrete tokens to automatically generate better prompts from a template.
Zhong et al. \cite{Zhong:2021:Optiprompt} developed OPTIPROMPT, which replaces AUTOPROMPT's discrete token search with a continuous vector search.
Qin et al \cite{Qin:2021:LearningHowToAsk} relax the constraints on continuous word embeddings to create ``soft prompts'' that avoid emphasizing misleading tokens such as gendered pronouns.
Elezar et al. \cite{Elazar:2021:ParaRel} created PARAREL, a benchmark for measuring the consistency of model predictions when prompts are paraphrased but the semantic meaning remains constant.
We extend this work using a visual analytics approach to qualitatively evaluating multiple prompts simultaneously for interpreting model performance.

\subsection{Visual Analytics For LLM Interpretability}
\label{sec:visualization_for_interpretability}
Visual analytics has become a popular approach for analyzing and interpreting machine learning models \cite{Hohman:2019:VAinDeepLearning}.
One method of interpreting model performance is directly visualizing the model's internal representations as a form of explanation \cite{Wang:2020:CNNExplainer}.
In deep learning models that perform natural language tasks such as Long Short Term Memory (LSTM) \cite{Hochreiter:1997:LSTM} and sequence-to-sequence (seq2seq) models \cite{Bahdanau:2014:seq2seq}, visualization techniques have been shown to help debug and explain how neural layers transform input sequences into final predictions \cite{Strobelt:2019:Seq2seqVis}.
Alternatively, visualizations can help users probe models from the outside by structuring the input process and supporting interactive analysis of model outputs.
VizSeq is a visual analysis toolkit for interactively evaluating LLM task benchmarks \cite{Wang:2019:VizSeq}.
Other tools present a visual analytics workflow to analyze changes in LLM weights under various task-specific scenarios \cite{Tenney:2020:LIT}.
These tools share a focus on human-in-the-loop workflows and interactivity to integrate valuable feedback during model validation \cite{Wang:2021:HITLNLP}, which \textit{KnowledgeVIS} makes heavy use of.

With transformers, new visual analytics approaches for interpreting how and why they work have been developed.
Several tools focus on visualizing the internal prediction process of transformers as inputs are fed through each layer of the model \cite{Vig:2019:BertViz, Hoover:2020:ExBERT, Derose:2020:AttentionFlows}.
In particular, Dodrio visualizes the connection between attention weights and linguistic knowledge such as syntactic dependencies \cite{Wang:2021:Dodrio}.
However, there is an active discussion as to whether attention weights in transformers can be used as a source of interpretation for model performance \cite{Clark:2019:AnalyzingBERTAttention, Jain:2019:AttentionIsNotExplanation, Atanasova:2020:ExplainingUsingWeights}.
Following our argument above, prompts can help users validate model performance by instead probing the model from the outside.
PromptIDE is a visualization interface for experimenting with prompt variations, visualizing prompt performance, and iteratively optimizing prompts \cite{Strobelt:2023:PromptIDE}.
LMDiff visually compares the difference in rank for all tokens in a single prompt between two different LLMs to facilitate comparison of model performance \cite{Strobelt:2021:LMdiff}.
We contribute to current research on human-in-the-loop workflows using prompts by introducing text visualization techniques for comparing multiple prompts simultaneously to validate model performance.

\section{Design Challenges and Goals}
\label{sec:design}

Our goal is to build a system for discovering learned associations from text representations in BERT-based language models.
\add{For natural language processing (NLP) researchers and engineers, interactively exploring model performance across different tasks can help build trust in models before deployment \cite{Wang:2021:HITLNLP}.}
Fill-in-the-blank sentences can help connect what large language models (LLMs) have learned with downstream tasks to interpret how they work \cite{Roberts:2020:KnowledgeLMParameters}.
Yet existing work in this space is primarily based around quantitatively evaluating prompts one at a time against an objective gold standard  \cite{Petroni:2019:LAMA, Jiang:2020:LPAQA, Shin:2020:Autoprompt, Qin:2021:LearningHowToAsk, Zhong:2021:Optiprompt}.
Our approach, a human-in-the-loop solution for qualitatively exploring multiple prompts simultaneously, overcomes several limitations by addressing the following key design challenges.

\subsection{Design Challenges}
\label{sec:design_challenges}

\noindent\textbf{C1  }
\textbf{Creating effective prompts.}
It remains an open challenge to explain how internal text representations in LLMs are translated into natural language understanding for completing tasks \cite{Roberts:2020:KnowledgeLMParameters}.
Using fill-in-the-blank sentences as prompts for LLMs formats queries intuitively as natural language tasks \cite{Liu:2023:PromptingSurvey}.
For example, Table~\ref{tab:prediction_data} shows how associations help demonstrate learned complex relationships.
A system should enable users to easily create, format, and test their own prompts.

\medskip
\noindent\textbf{C2  }
\textbf{Testing multiple prompts at once.}
A single prompt can give limited understanding of what association LLMs are making based on the context of the sentence \cite{Warstadt:2019:AnalysisBERTKnowledge, Elazar:2021:ParaRel, Rogers:2020:Bertology}.
To enable more effective prompt testing, we can evaluate multiple prompts simultaneously that isolate dependent variables, such as the bold subjects in Table~\ref{tab:prediction_data}.
Our interface should encourage users to test multiple prompts through intuitive input design.

\medskip
\noindent\textbf{C3  }
\textbf{Probing different types of relationships.}
While fill-in-the-blank prompts can be designed to elicit a single fact or relationship \cite{Petroni:2019:LAMA}, LLMs also learn multiple correct responses to the same prompt, subjective answers, and associations such as stereotypes and domain-specific knowledge \cite{Rogers:2020:Bertology}.
Table~\ref{tab:prediction_data} shows how subjective, open-ended prompts can elicit meaningful predictions that help users interpret model performance.
Providing several prompt examples can guide users to more effectively design a wide variety of probes for eliciting different relationships.

\medskip
\noindent\textbf{C4  }
\textbf{Finding insights in a large search space.}
As shown in Table~\ref{tab:prediction_data}, predicted tokens and their probability scores present several analytical challenges.
LLMs often have large vocabularies that lack useful stratification \cite{Devlin:2019:BERT}.
Tokens themselves can be both specific to a prompt and generalizable across prompts, duplicated across prompts, and/or unique within a prompt.
Methods for grouping, filtering, searching, and arranging predictions and their probabilities can aid users in discovering insights.

\begin{figure*}[!t]
  \centering
  \setlength{\abovecaptionskip}{0pt}
  \includegraphics[width=\linewidth]{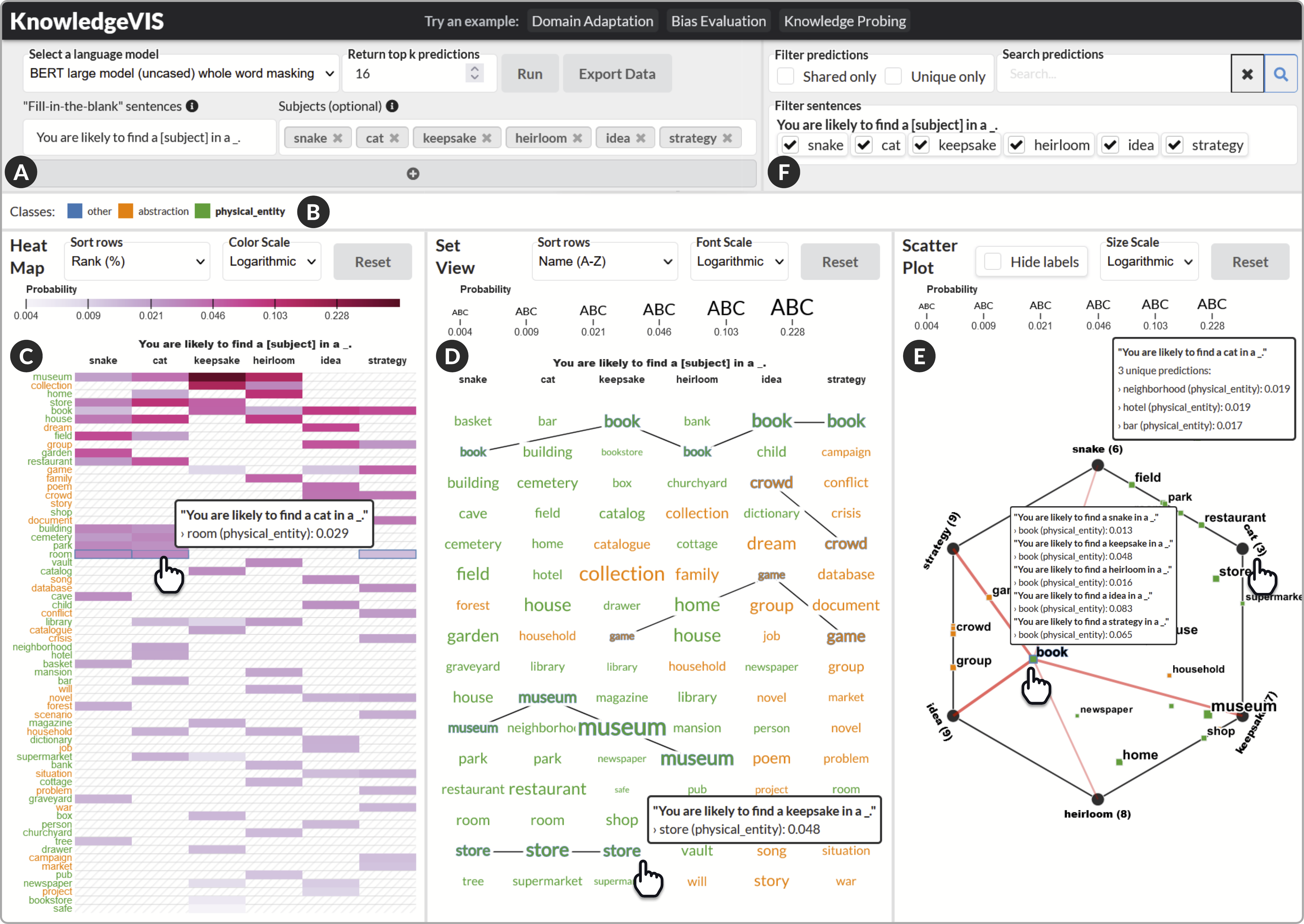}
  \caption{%
    \textit{KnowledgeVIS} integrates multiple views to reveal associations that LLMs learn during training.
    Above, a user investigates whether BERT exhibits associations that reveal learned conceptual relationships (Sect.~\ref{sec:case_knowledge}), helping them interpret how BERT works.
  }%
  \label{fig:interface}
  \vspace{-1em}
\end{figure*}

\subsection{Design Goals}
\label{sec:design_goals}
We developed design goals that address the key challenges raised in Sect.~\ref{sec:design_challenges} and align with our interface components:

\medskip
\noindent\textbf{G1  }
\textbf{Intuitive visual interfaces for structuring prompting.}
Clearly communicating how to input prompts through visual design can help guide users to more rapidly test for learned associations.
``Fill-in-the-blank'' prompt inputs should be open-ended to encourage creativity (\textbf{C1}) with flexible formatting rules to enable rapid generation of different prompt variations (\textbf{C3}).
At the same time, arranging inputs to facilitate comparison across prompt variations should encourage more thoughtful probing (\textbf{C2}).

\medskip
\noindent\textbf{G2  }
\textbf{Useful grouping of prompts and predictions.}
Providing additional structure to several large sets of predictions can help reduce their complexity (\textbf{C4}).
We aim to differentiate predictions by their semantic relatedness and communicate this distinction visually while respecting the input structure of prompts.
Highlighting new connections could enable domain experts to use their own knowledge for evaluating predictions more effectively (\textbf{C2}, \textbf{C3}).

\medskip
\noindent\textbf{G3  }
\textbf{Expressive and interactive views for discovering insights.}
Fluid transitions between levels of abstraction in the data can surface patterns across multiple sets of predictions that connect learned associations to model performance (\textbf{C4}).
We aim to support several low-level tasks including identifying highly salient or unique predictions, comparing both individual and sets of unique and shared predictions between prompt variations, and summarizing groups of predictions as over/under performing subsets for further investigation.
Incorporating these tasks across several coordinated views can help users connect these low-level tasks with high-level model performance (\textbf{C2}, \textbf{C3}).

\section{KnowledgeVIS}
\label{sec:system}

Based on the design challenges and goals in Sect.~\ref{sec:design}, we developed \textit{KnowledgeVIS} (Fig.~\ref{fig:interface}), a human-in-the-loop visual analytics system for comparing fill-in-the-blank prompts to uncover associations from learned text representations.
Our interface tightly integrates multiple coordinated views with a novel prediction clustering technique.
Users structure the prompt generation and testing process using the \textit{Prompt Interface} (Sect.~\ref{sec:prompt_interface}).
Then, the \textit{Predictions View} (Sect.~\ref{sec:predictions_view}) visualizes predictions at multiple levels of abstraction across a \textit{Heat Map} (Sect.~\ref{sec:heat_map}), \textit{Set View} (Sect.~\ref{sec:set_view}), and \textit{Scatter Plot} (Sect.~\ref{sec:scatter_plot}).
Finally, the \textit{Filters Panel} (Sect.~\ref{sec:filters_panel}) allows users to fine-tune their analysis.
Throughout, we describe how each design goal (\textbf{G1-3}) is addressed in our interface.

\subsection{Prompt Interface}
\label{sec:prompt_interface}
The \textit{Prompt Interface} (Fig.~\ref{fig:interface}A) guides users to create multiple prompts using a grid structure and intuitive text formatting that promote structured exploration of different prompt variations.
Users start here to begin exploring the capabilities of prompting for eliciting associations.

To guide users in eliciting interesting sets of predictions, we structure prompt inputs in several ways.
To help users probe for different types of relationships, we provide buttons that load examples related to domain adaptation, bias evaluation, and knowledge probing (\textbf{G1}).
We explore these examples in depth in Sect.~\ref{sec:cases_evaluation}.
Prompts are written in open-ended text inputs aligned in a 2-column grid (\textbf{G1}).
The horizontal dimension promotes comparison across paraphrased sentences.
Users write prompt ``templates'' in the left column that must include a single underscore character for the LLM to fill in.
Users can then use the right column to input any number of ``subjects'', to be filled in to the template prompt using an additional \textbf{[subjects]} mask anywhere in the prompt.
This allows users to intuitively create single- or multi-token variations on the given template prompt, i.e., paraphrasing prompts grammatically.
The vertical dimension provides additional rows for writing template/subject pairs to compare semantically similar prompts that may differ grammatically.
Finally, we mirror the grid structure hierarchically by template $\rightarrow$ subject (\textbf{G2}) in the \textit{Predictions View}, described in Sect.~\ref{sec:predictions_view}, helping users reflect on how they can refine their prompts.

Prompts are evaluated in real-time using an API that interfaces with a Python Flask server running PyTorch implementations of the LLMs.
We use the HuggingFace Transformers \cite{Wolf:2020:HuggingfaceTransformers} API to load models and perform masked word prediction with the user-created prompts.
Users can select different pre-trained masked language models from a drop-down list to compare performance and add new models easily through the HuggingFace Transformers model library.
To reduce the complexity of the prediction space, we also let users choose the top $k$ tokens to retrieve and visualize.
Finally, users can export the extracted data in the format of Table~\ref{tab:prediction_data} for further investigation.

\subsection{Predictions View}
\label{sec:predictions_view}
We provide several facilities in the \textit{Predictions View} (Fig.~\ref{fig:interface}B-E) to promote exploration and discovery of insights about what a model has learned.
First, we group predictions by semantic similarity using a novel taxonomy-based clustering technique developed in Sect.~\ref{sec:clustering_predictions}.
Then, we visualize the prompt, prediction, and cluster data as shown in Table~\ref{tab:prediction_data} at multiple levels of abstraction across three interactive plots.
Each plot provides unique advantages for different analysis tasks that the other plots do not, that together help users find patterns better than a single plot could.
In Sect.~\ref{sec:cases_evaluation} we describe how the plots are used in tandem in each use case as well as in the expert evaluation to help guide participants' analysis process and thinking.

The \textit{Heat Map} (Sect.~\ref{sec:heat_map}) makes it easy to accurately identify and compare individual probabilities across prompt variations (columns) and semantic clusters (rows).
The grid structure uniquely highlights\strike{ missing} words \add{not shared} between prompts to help users find outliers.\strike{ You cannot do this easily in the \textit{Set View} or \textit{Scatter Plot}.}
The \textit{Set View} (Sect.~\ref{sec:set_view}), inspired by parallel tag clouds \cite{Collins:2009:ParallelTagClouds}, facilitates in-depth comparison of word sets across multiple prompts, as well as rank order analysis.
Selecting a predicted word aligns each occurrence across prompts along a common baseline and shows a novel selected word rank order view.\strike{ You cannot do this easily in the \textit{Heat Map} or \textit{Scatter Plot}.}
The \textit{Scatter Plot} (Sect.~\ref{sec:scatter_plot}) projects predictions in a low-dimension space and uses a dust-and-magnet metaphor \cite{Soo:2005:DustAndMagnet} to position prompts as points of interest, revealing new relationships between predictions at the data set level.
For example, common predictions more relevant to a subset of prompts, as well as unique predictions sharing a relationship between two prompts, are visually grouped.\strike{ You cannot do this easily in the \textit{Heat Map} or \textit{Set View}.}

\subsubsection{Clustering Predictions}
\label{sec:clustering_predictions}
Users may not have an intuitive sense for seeing patterns in shared meaning across predicted tokens, especially as the number of predicted tokens grows.
This meta-information can be useful to determine, e.g., specific training biases towards higher-level concepts that may or may not be relevant to the semantic meaning of the prompt.
To help users discover patterns in prediction sets, we aim to \add{group and describe semantically similar predictions}\strike{ automatically find and label sets of predictions based on shared semantic similarity}.
\add{While topic modelling has been used to assign manually-sourced candidate labels to lists of terms based on the co-occurrence of labels and terms \cite{Lau:2011:AutomaticLabeling}, there are a lack of methods for automatically discovering appropriate labels without frequency data.}
\add{To overcome this,} our solution \add{uses a hierarchical taxonomy of word sense, or the meanings of words, to} algorithmically \add{generate and label clusters of words based on their shared semantic meaning (\textbf{G2}). We color tokens by their cluster label (\textbf{G3}) in each visualization (Fig.~\ref{fig:interface}C-E).}\strike{ generates clusters and allows the user to investigate them across each of the plots in the \textit{Predictions View} (\textbf{G2}).}

\begin{enumerate}[topsep=4pt, itemsep=4pt]
  \item Compute a distance matrix of pairwise Wu-Palmer similarities \cite{Wu:1994:WuPalmerSimilarity} between all unique predicted words. Compared with other popular measures such as word vector similarity \cite{Mikolov:2013:word2vec}, Wu-Palmer led to better cluster labels downstream.
  \item Perform hierarchical clustering \cite{Mullner:2011:HierarchicalClustering} over the distance matrix using Ward linkage \cite{Ward:1963:WardsMethod}. Affinity propagation \cite{Frey:2007:AffinityPropagation}, while often used with similarity-based distance metrics, requires unintuitive manual tuning of the damping factor to avoid overfitting.
  \item Determine the optimal number of clusters based on either the maximum silhouette coefficient \cite{Rousseeuw:1987:SilhouetteScore} or a user-defined cut-off threshold on hierarchical clusters.
  \item Automatically label each cluster by computing the lowest common hypernym (LCH) between all words in the set using WordNet \cite{Miller:1995:WordNet}. Hypernyms (words with a broad meaning that more specific words fall under such as ``color'' and ``red'') are an approximate yet informative label for a majority of the open-class word predictions returned by LLMs such as nouns, verbs, etc.
\end{enumerate}

\strike{The results of a sample run are shown in Fig.~\ref{fig:interface}B-E. Our algorithm successfully clustered 64 unique words into 3 groups based on shared semantic similarity labeled by lowest common hypernym. We found a user-defined threshold of a maximum of 15 clusters tended to produce semantically relevant labels. If the silhouette coefficient determines an optimal number of clusters greater than 15, we return all predicted tokens with a single cluster (``other'') to avoid misrepresenting the data. We visualize these clusters in each of the following plots by coloring tokens according to their cluster label (\textbf{G3}), as shown in Fig.~\ref{fig:interface}C-E.}

\subsubsection{Heat Map}
\label{sec:heat_map}
The \textit{Heat Map} (Fig.~\ref{fig:interface}C) plots a uniform grid of all unique predicted tokens vertically as rows and all prompts as columns.
Each cell is colored by the probability of the given word (row) occurring in the prompt (column).
Much like a data table, visualizing predictions in this way allows users to quickly identify and compare small details in the entire data set at a glance, such as individual probabilities, cluster labels, and how they compare across prompts (\textbf{G3}).

Several design considerations enhance analysis capabilities.
If a given word (row) is not in the set of predictions returned for a given prompt (column), that cell is left unshaded, and a crosshatch pattern is used to differentiate missing values from the background of the interface.
We provide row lines to aid in visually scanning across columns with unshaded cells.
We found that probabilities tended to range non-linearly, with very few high-probability words relative to many low-probability words.
Thus, for broad comparison across all predicted words, we chose as default a logarithmic color ramp from light to dark pink applied to the global extents of the range of probabilities; i.e., the darkest shade of pink is the highest probability in the data set, and the lightest shade of pink is the lowest.
We provide a legend and a drop-down list to select between the default logarithmic scale as well as a linear scale, helping the user more accurately compare probabilities.
For example, when comparing a small subset of probabilities for a single prompt, a linear scale is more appropriate.
To connect the structure of the inputs with output predictions, we label the columns of the \textit{Heat Map} hierarchically (\textbf{G2}).
We nest subjects as column names below sentence templates that span several columns, to indicate group membership.
If no subjects are put in, i.e., the template is the only prompt to evaluate, then we treat the template as a subject and only show the template in the lowest level of the hierarchy.

We also implemented interactions that provide additional details on demand.
Hovering over a cell displays a tooltip showing the prompt (column), prediction (row), cluster, and probability.
Rows can be sorted by word top-down in one of four ways: (1) alphabetically; (2) rank order (i.e., by probability of occurring in a prompt, from left to right); (3) grouped by cluster, alphabetically; and (4) grouped by cluster, rank order.
Groups are sorted top-down alphabetically by cluster label.

\begin{figure}[tb]
  \centering
  \setlength{\abovecaptionskip}{0pt}
  \includegraphics[width=\columnwidth]{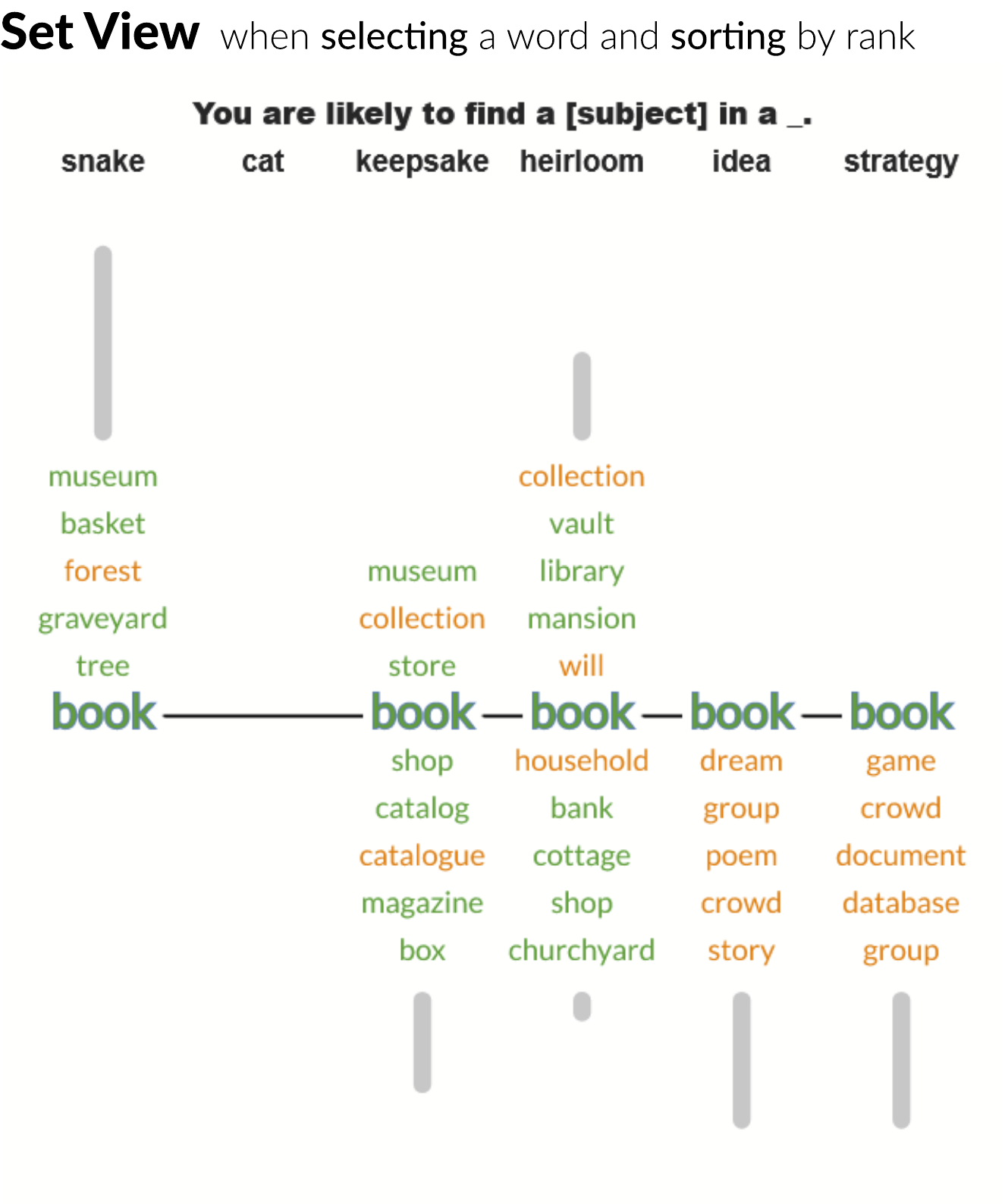}
  \caption{%
  The \textit{Set View} showing our variation on the parallel tag cloud layout, a stepwise degree of interest list based on fisheye menus \cite{Bederson:2000:FisheyeMenus} for a selected word when sorting by rank, described in Sect.~\ref{sec:set_view}.
  }%
  \label{fig:set_view_ranked}
  \vspace{-1em}
\end{figure}

\subsubsection{Set View}
\label{sec:set_view}
The \textit{Set View} (Fig.~\ref{fig:interface}D) arranges sets of the top $k$ predicted tokens returned for each prompt into parallel columns.
Similar to cell color in the \textit{Heat Map}, the font size for each word (row) is scaled by the probability of occurring in the prompt (column).
Inspired by parallel tag clouds \cite{Collins:2009:ParallelTagClouds}, this plot \add{shows the degree of overlap between prompts (columns) by drawing edges between shared tokens, making shared predictions more visually salient at a glance}\strike{ combines the analytical capabilities of parallel coordinates with the text visualization capabilities of tag clouds}.
\add{To overcome difficulties in interpreting continuous word probabilities, such as the cell shading in the \textit{Heat Map}, users can also sort columns by rank, encoding probability instead on a discrete scale and making differences between prompt variations more interpretable (\textbf{G3}).}\strike{ The \textit{Set View} helps users easily identify highly probable words across prompts, compare sets of shared and unique predictions between prompt variations, and perform in-depth rank order analysis (\textbf{G3}).}

To help users see new patterns in the data, we implemented hover and select interactions that change the arrangement of the words.
Hovering on and off a word will temporarily draw a connector line from the word to all other occurrences of that word in other columns, while hiding the connector line if it crosses a column that does not contain the hovered word.
While hovering, selecting the word will shift the columns containing other occurrences of that word to align the words along the same horizontal baseline.
The connector line will then remain drawn even after hovering off.
Selecting other words will align the columns along a new baseline and automatically adjust the previously drawn connector lines.
Deselecting a selected word will remove the connector line and reset the column alignment.
If a word does not occur in one or several columns when it is selected, each of those columns is set to a lower opacity to more clearly show a lack of membership of the word in that set.
Finally, selecting a word while sorting by rank order transitions the view to a stepwise degree of interest list (Fig.~\ref{fig:set_view_ranked}), similar to fisheye menus \cite{Bederson:2000:FisheyeMenus}.
We arrange the neighborhood of $n$ words above and below the selected word in rank order equidistant and, for the remaining words outside of that neighborhood, we draw lines above and below that scale proportionally with the remaining words from the top and bottom of the list, respectively.
Users can accurately compare line heights along the same baseline to determine the difference in rankings between columns (\textbf{G3}).
\add{Words not occurring in a column in this view are set to zero opacity since no line(s) will be drawn for that column.}

As in the \textit{Heat Map}, we label columns hierarchically (\textbf{G2}), provide a legend and drop-down list for logarithmic and linear font scales, draw tooltips when hovering over words (showing the prompt, prediction, cluster, and probability), and let the user sort words top-down (alphabetically, rank order, and grouped by cluster).
Collins et al. found that ordering words alphabetically offers two main advantages over rank order: (1) it saves horizontal space, since the largest words are less likely to occur next to each other; and (2) it helps users visually scan for words of interest \cite{Collins:2009:ParallelTagClouds}.
Thus, we make this the default sorting option for rows.

\subsubsection{Scatter Plot}
\label{sec:scatter_plot}
The \textit{Scatter Plot} (Fig.~\ref{fig:interface}E) positions predicted tokens as vectors based on their probability of occurring for each prompt or ``point of interest'' (POI) in a 2D coordinate space, using a layout technique derived by Olsen et al. \cite{Olsen:1993:VIBE}.
\add{%
Predictions closer to POIs (prompts) indicate a higher probability of occurring for that prompt.
Groups of predictions in between POIs reveal a unique relationship between prompts that cannot be seen in the other two plots.
Predictions that occupied the shared axis of two POIs revealed unique prompt interactions that aligned with semantic cluster labels (\textbf{G2}).
}%
Because this approach reduces the dimensionality of predictions, we allow users to drag POIs, similar to a dust-and-magnet metaphor \cite{Soo:2005:DustAndMagnet}, to create new arrangements of data marks and avoid visual artifacts based on the initial layout.
Overall, the process of arranging predictions spatially relative to prompts can help users uncover new patterns and related predictions at the data set level (\textbf{G3}).

To help users read the \textit{Scatter Plot}, we provide several visual embellishments.
Data marks are labeled with the predicted word or prompt they represent; we implemented occlusion to show only the top label when several labels overlap.
Users can hide these labels with a checkbox.
For three POIs, we also draw a differently colored background for the bounded region containing points most closely associated with each POI.
When predicted words are unique to a single prompt (POI), the layout algorithm avoids plotting them all at the same position of the POI.
Instead, we collect the unique predictions and display each word, cluster, and probability set in a tooltip when hovering over a POI.
We append a count of the number of unique predictions for each prompt to every POI label.
\add{%
Finally, to visually distinguish relationships between adjacent points of interest, we draw the convex hull around all POIs.
Dragging a POI dynamically adjusts the convex hull, to show a shared axis or axes with adjacent neighbors.
}%

\strike{The initial layout of predicted words and POIs can create coincidental visual patterns that may not be present in the data. Users may also want to create new shared axes between POIs that may not be apparent. We solve this using a dust-and-magnet approach \cite{Soo:2005:DustAndMagnet}, allowing users to freely drag POIs around the plot. The layout algorithm dynamically updates the position of the dragged POI and determines the new location for each prediction.}

Because the plot reduces the dimensionality of each predicted word, we lose information contained in the original vector, i.e., the probabilities of the predicted word occurring in each prompt.
We provide three solutions to recover this information.
The first is a details-on-demand tooltip when hovering over a predicted word that lists non-zero probabilities of the predicted word occurring for each prompt.
The second is encoding the maximum probability of a predicted word occurring for all prompts as height/width of the \textit{Scatter Plot} point, as well as font size of the label, on either a logarithmic or linear scale, similar to the \textit{Heat Map} and \textit{Set View}, and providing a legend.
The third is drawing lines on hover from the predicted word to each of the non-zero probability prompts (Fig.~\ref{fig:interface}E), double-encoding stroke width and opacity to each line using the same scale.
For example, a weak relationship (low probability) between a POI and predicted word is shown with a smaller point and label and, when hovering, with a faded-out thin line, whereas a strong relationship (high probability) is shown with a larger point, label, and vibrant thick line.

\subsection{Filters Panel}
\label{sec:filters_panel}
With the \textit{Filters Panel} (Fig.~\ref{fig:interface}F), users can filter prompts by directly toggling subjects nested under their template sentence.
Arranging the prompt filter using the template $\rightarrow$ subject hierarchy (\textbf{G2}) can promote a wider variety of prompt testing, e.g., by subsetting prediction sets with specific semantic relationships and comparing the results against other subsets.
Within the currently visible subset of prompts and predictions, we provide two additional global prediction filter operations:  ``shared only'' and ``unique only'' checkboxes.
The ``shared only'' checkbox filters predictions that are shared between all of the currently visible prompts.
This removes all missing cells in the \textit{Heat Map} and aligns the words along the same horizontal baseline in the \textit{Set View}, helping users more easily find common predictions and quickly compare relative probabilities.
Similarly, the ``unique only'' checkbox filters predictions that are unique to each visible prompt.
This reduces the rows of the \textit{Heat Map} so that each contains a single shaded cell, and removes all points from the \textit{Scatter Plot}, as none share relationships with the other prompts.
The \textit{Set View} can help compare biases in each unique set of predictions across prompts, while sorting the \textit{Heat Map} by cluster can reveal the distribution of classes of words, such as classes unique to a specific prompt.
Users can also use the provided search box to highlight all instances of a specific prediction word across all plots.

\section{Use Cases and Expert Evaluation}
\label{sec:cases_evaluation}

\begin{figure*}[!t]
  \centering
  \setlength{\abovecaptionskip}{0pt}
  \includegraphics[width=\linewidth]{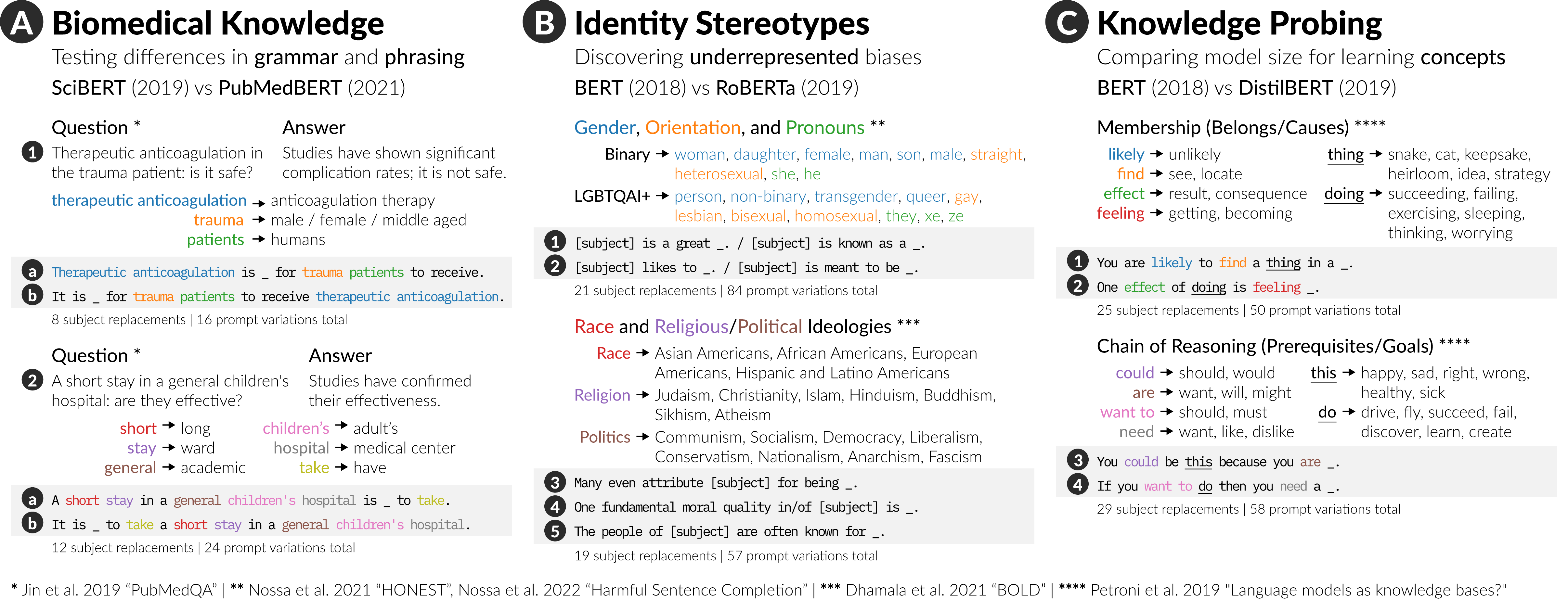}
  \caption{%
    Data sets and prompts used in our use cases (Sect.~\ref{sec:cases_evaluation}) to show how \textit{KnowledgeVIS} can help NLP researchers and engineers interpret LLMs.
  }%
  \label{fig:use_cases}
  \vspace{-1em}
\end{figure*}

In this section, we demonstrate the capabilities of \textit{KnowledgeVIS} for prompt engineering and immediate visual analysis of fill-in-the-blank sentence predictions with three use cases (Fig.~\ref{fig:use_cases}) and an expert evaluation.
Our use cases comprised 114 subject replacements across 15 fill-in-the-blank sentences, totaling 289 prompt variations to evaluate.
\strike{We demonstrate in each use case how the \textit{Prompt Interface}, \textit{Predictions View} and \textit{Filters Panel} enabled rapid generation, comparison, and filtering. Additionally, }The use cases were designed for NLP researchers and engineers who are \add{increasingly using LLMs as ``black boxes'' for downstream tasks}\strike{ often not experts in the underlying technology of the language model and typically use models for downstream applications} \cite{Hohman:2019:VAinDeepLearning} such as discourse analysis and classification.
Compared to automated methods using quantitative metrics determined \textit{a priori}, \textit{KnowledgeVIS} leverages human intuition and domain expertise to guide LLM evaluation \cite{Wang:2021:HITLNLP}.
This enabled experts to suggest new ways to adapt and improve LLMs in their own research and applications, towards ``closing the NLP loop'' in model development (Sect.~\ref{sec:closing_the_loop}).

\add{%
The results are based on interpreting word probabilities using visual encodings.
This presents analytical trade-offs when choosing scales (linear and log) as well as encodings (color, font and marker size).
For example, log scale tended to more prominently show patterns, but can be misleading.
Patterns based on positional encoding in the \textit{Scatter Plot} are also more likely to draw a user's attention than size encodings.
To address this, we detail our method for generating and evaluating prompts in the second paragraph of each use case.
We acknowledge our qualitative approach may be subject to pre-attentive biases.
}%

First, we tested how grammar and phrasing affected domain-specific LLMs when replicating expert human answers is required.
We modified a biomedical question-answer data set, PubMedQA \cite{Jin:2019:PubMedQA}, by formatting yes/no/maybe questions as fill-in-the-blank sentences, then queried SciBERT \cite{Beltagy:2019:SciBERT} and PubMedBERT \cite{Gu:2021:PubMedBERT}, both pre-trained on large unlabeled scientific document corpora.

Second, current LLMs are regularly fine-tuned on pre-trained general-domain LLMs such as BERT \cite{Devlin:2019:BERT} and RoBERTa \cite{Liu:2019:RoBERTa} and inherit well known stereotypical associations such as gender bias.
Yet automated auditing systems requiring social categories and quantitative metrics to be predetermined are prone to missing underrepresented stereotypes \cite{Bender:2021:StochasticParrots}.
We discovered contextualized gender, orientation, pronoun, race, religious, and political stereotypes between BERT and RoBERTa using subsets of the HONEST \cite{Nozza:2021:HONEST, Nozza:2022:HONEST-LGBTQIA+} and BOLD \cite{Dhamala:2021:BOLD} data sets.

Third, as LLMs increase in size, it is useful to understand limitations at different model scales on whether associations not explicitly trained for such as membership (belongs, causes) and chain of reasoning (goals, prerequisites) are learned.
We compared complex learned concepts between the large-scale BERT \cite{Devlin:2019:BERT} and small-scale DistilBERT \cite{Sanh:2019:DistilBERT} models based on the LAMA knowledge probe \cite{Petroni:2019:LAMA}.

Finally, we conducted an expert evaluation with six academic NLP researchers and engineers.
The experts generated new examples to uncover insights, including: (1) a lack of understanding for semantic roles in all three general-domain models; (2) unexpected biases towards less common words in general-domain models; and (3) differences in grammar robustness between domain-specific models.

\subsection{Use Case: Biomedical Knowledge}
\label{sec:case_biomedical}
How do domain-specific models compare based on robustness to grammar and phrasing when expert human answers are expected?
In Fig.~\ref{fig:use_cases}A, we evaluated two PubMedQA \cite{Jin:2019:PubMedQA} questions \textbf{(1 and 2)} by formatting two grammatically different but semantically similar prompts per question \textbf{(1a/b and 2a/b)} and using the \textit{Prompt Interface} to replace key phrases.

We mostly \add{replaced subjects in multiple locations within a sentence with a single word (e.g., ``short'' replaced by ``long'')}\strike{ performed single subject replacements} in the \textit{Prompt Interface}.
The \textit{Heat Map} was critical for finding predictions not shared between prompts.
The \textit{Set View} was also useful for understanding ranking patterns that the \textit{Heat Map} cannot represent, such as when different prompts exhibit similar prediction sets but differently ordered.
The logarithmic scale was mostly used, as the models generally returned few highly probable predictions.

For PubMedBERT, key phrases and synonyms changed recommendations regardless of the context of the sentence.
For example, the \textit{Heat Map} showed missing entries between therapeutic anticoagulation (``ideal'', ``significant'', ``imperative'', ``standard'') and anticoagulation therapy (``costly'', ``expensive'', ``complex''), while the \textit{Set View} showed some positive associations for ``patients'' and not for ``humans''.
PubMedBERT also associated ``long'' with sets of words including ``expensive'', ``dangerous'' and ``bad'', while ``short'' was ``safe'', ``convenient'', ``simple''. 
This association persists even when other phrases change, raising concerns that the model is not recognizing important syntax and only focusing on ``short''/``long'', which may be a consequence of its training on PubMed articles.
Thus, PubMedBERT can be good for general-purpose recommendations where grammar is consistent with the training data.
Where key phrases change or are important to consider, it may be hard for PubMedBERT to unlearn certain strong associations.

For SciBERT, key phrase changes are generally ignored while the context and grammar of the sentence more heavily change predictions than with PubMedBERT.
For example, where SciBERT is consistent (i.e. most rows filled in the \textit{Heat Map}) across subject replacements for \textbf{1a}, it is similarly consistent yet opposite for \textbf{1b} (``recommended'' vs ``not'').
Similarly, SciBERT finishes the sentence for \textbf{2a} with ``difficult'', ``easy'', ``hard'', and ``simple'' across almost all variations, not making any recommendation, while using ``take'' in \textbf{2b} results in recommendations like ``able'', ``possible'', ``required'' and ``likely''.
Sentence phrasing affects how much SciBERT recommends something.
This could make SciBERT good for learning associations on key phrases, but harder to adapt to new grammars and phrasings.

Overall, while both models are susceptible to grammar and phrasing issues, PubMedBERT tends to associate recommendations with certain key phrases, whereas SciBERT bases it on the word order.

\subsection{Use Case: Identity Stereotypes}
\label{sec:case_stereotypes}
How can important yet underrepresented identity stereotypes be discovered in general-purpose LLMs?
In Fig.~\ref{fig:use_cases}B, we modified prompts from the HONEST \cite{Nozza:2021:HONEST, Nozza:2022:HONEST-LGBTQIA+} data set \textbf{(1 and 2)} for measuring gender, orientation, and pronoun biases between binary and LGBTQIA+ communities.
We also modified prompts from the BOLD \cite{Dhamala:2021:BOLD} data set \textbf{(3, 4, and 5)} to further investigate United States racial, religious, and political identity stereotypes.
Importantly, this use case only highlights a subset of identities and doesn't account for intersectionality \cite{Crenshaw:1989:Intersectionality} (i.e., identifying with multiple groups); future work should investigate intersectional identity completions in masked language models.

\strike{We primarily leveraged multiple subject replacements, }\add{In contrast to the previous use case, we primarily replaced subjects in a single location within a sentence with multiple words (e.g., ``woman'', ``daughter'', etc.) and used} rank-order views and set membership in the \textit{Set View} and \textit{Scatter Plot}.
Using a linear scale, we saw few predictions more likely than others and high probability, making ranking an important metric to distinguish behavior.
We used the \textit{Set View} and \textit{Scatter Plot} to surface shared sets of predictions and find unique predictions that stand out, e.g., specific predictions unique to a single set, or a prediction that is higher for a prompt than the rest.
The shared/unique filters were critical for reducing the prediction space in the \textit{Set View}, such as when few/many predictions are shared and unique predictions reveal shared prompt groups.

\medskip
\noindent\textbf{Gender, orientation, and pronouns. }
For BERT, as expected, binary labels were more often associated with gender norms and positivity compared with LGBTQIA+ labels being misclassified with stereotypical and negative exceptions (e.g., ``beautiful'' and ``admired'', versus ``different'' and ``temporary'').
For RoBERTa, we found bias in associations with morality and gender norms to be less frequent and isolated overall.
Yet one unexpected and concerning association in both models (i.e. high ranking and shared relationships in the \textit{Set View} and \textit{Scatter Plot}) is ``lesbians'', ``women'', and ``female'' with many LGBTQIA+ labels.
Could more LGBTQIA+ content be associated with or written for women, or is this a more ingrained misclassification bias?
Why is this association not debiased in RoBERTa given its performance with other labels?
Further, the association of LGBTQIA+ labels and morality, particularly themes observed such as ``evil'' and ``sin'', is then more likely to refer to women than men.
It is unclear in what direction the association between woman, LGBTQIA+ labels, and morality is targeted. 
Another unexpected association was made between men, sports and sexuality in RoBERTa. 
Interestingly, the inclusion of heterosexual/homosexual labels reveals associations with ``athletes'', ``coaches'', ``players'', and ``leaders''. 
This could reveal a mechanism of the training surfacing to debias identity labels missing a particular association. 
Finally, we observed difficulties in both models with understanding LGBTQIA+ pronouns at all, which could point to an unintentional bias with little training data to support masked language modeling.

\medskip
\noindent\textbf{Race, religion, and politics. }
For BERT, given its poor performance on gender, orientation, and pronoun labels, we were surprised to find very few negative associations with race, religion, and politics overall.
Yet we found Hispanic and Latino Americans had very few unique associations (i.e. low probability and variability in the \textit{Set View}) compared with other labels; it is likely that Hispanic and Latino Americans are underrepresented in BERT's training.
For RoBERTa, we found a higher rate of bias and negative associations (i.e. high probability and variability in the \textit{Set View}) across underrepresented groups in the United States, such as Asian Americans with ``bullying'', ``discrimination''; Hispanic and Latino Americans with ``gangs'', ``homelessness''; and African Americans with ``slavery'', ``hardship''.
RoBERTa also exhibited strong biases in attributes, moral qualities, and affiliations between two different groups of religions.
In the \textit{Scatter Plot}, Islam/Hinduism/Christianity shared many points associated with marriage and morality compared to Judaism/Buddhism/Sikhism with peace and service (``polygamy''/``patriarchy''/``evil''/``oppressive'' vs ``tolerant''/``strong''/``good''/``compassion'').
It is surprising that RoBERTa has such ingrained stereotypes given the debiasing shown in juxtaposition with otherwise stereotypical predictions from BERT. 
Interestingly, moral qualities of political ideologies were divided in RoBERTa into shared qualities between groups.
Using the \textit{Scatter Plot}, we identified associations along shared edges (Anarchism, Facism and ``violence''; Facism, Communism and ``weak'', ``evil''; Fascism, Conservatism and ``arrogance''; Nationalism, Conservatism and ``loyalty''; Conservatism, Liberalism and ``tolerance'', ``moderate'').
RoBERTa also frequently suggested Communism is ``Jewish'' (i.e. most rows filled in the \textit{Heat Map}), relating two identities and suggesting learned intersectional biases may exist, though overlapping identity associations were otherwise rarely seen in both models.

\subsection{Use Case: Knowledge Probing}
\label{sec:case_knowledge}
How well do LLMs learn complex relationships at different model scales?
In Fig.~\ref{fig:use_cases}C, we created templates to elicit associations with open-ended conceptual prompts and various subject replacements that test reasoning capabilities for membership \textbf{(1 - belongs and 2 - causes)} and chain of reasoning \textbf{(3 - prerequisites and 4 - goals)}, based on the LAMA \cite{Petroni:2019:LAMA} knowledge probe.

We \add{replaced subjects in multiple locations within a sentence with multiple words (e.g., ``effect'' and ``feeling'' in the same sentence) and} used the \textit{Scatter Plot} to observe higher-level patterns across prompts, helping us identify clusters of labels.
Some labels sit on the line between two subjects, which was very interesting.
Prediction clusters also provided evidence of learned semantic understanding of knowledge where subject associations matched their prediction hypernym labels (e.g., snake/cat produced mostly ``physical entity'' predictions, while strategy/idea produced mostly ``abstraction'' predictions).
We also used the \textit{Set View} selected word rank view (Fig.~\ref{fig:set_view_ranked}) to see how predictions compare across prompts, when some are higher/lower, lists that do not match, lists that are offset slightly, etc.

For BERT, associations are mostly unique and relevant to the subject replacements across all prompts.
Membership is highly dependent on the subject (i.e. the unique filter shows most rows in both the \textit{Heat Map} and \textit{Set View}).
For example, we saw differences between where you find, locate and see things (e.g., ``drawer'' vs ``building'' vs ``dream'', respectively), or when feeling, getting or becoming (e.g., ``satisfied'' vs ``older'' vs ``greater'', respectively).
Relationships can also be positive or negative -- consequence produces negative associations while result/effect share common positive associations (e.g., ``powerless''/``bad'' and ``good''/``desired'', respectively). 
BERT also understands conceptual pairs (i.e. groups of predictions sharing an edge in the \textit{Scatter Plot}). 
Snake/cat are animals found in a ``park'' or ``garden''; heirloom/keepsake are objects in a ``museum'' or ``collection''; a strategy/idea are found in a ``story'' or ``job''. 
The predictions followed their subject hypernym clusters (e.g., snake/cat produced mostly ``physical entity'' predictions, strategy/idea produced mostly ``abstraction'' predictions, while keepsake/heirloom were mixed). 
Chain of reasoning prompts produced similar results.
BERT showed unique and relevant prerequisites (i.e. few shared connector lines in the \textit{Set View}) for healthy/sick such as ``hungry'', ``tired'' or ``pregnant'', happy/sad such as ``alone'' or ``here'' and right/wrong such as ``stubborn'', ``smart'' or ``blind''.
Top predictions for goals such as drive and fly are correct (e.g., ``car''/``map'' and ``pilot''/``wings'', respectively).
Succeed and fail are opposite (``plan''/``strategy'' vs ``distraction''/``reason''), while discover, learn, and create are all associated with ``teachers'', ``lessons'', and ``partners''. 
BERT demonstrates a strong understanding of complex reasoning across both membership and chain of reasoning examples.

For DistilBERT, we found strong performance similar to BERT in making associations for belongs and goals, such as similar pair associations between snake/cat, drive/fly and discover/learn/create in the \textit{Scatter Plot}. 
We noticed associations were mostly noun-based and did not follow our scheme of separating membership and chain-of-reasoning. 
However, DistilBERT failed to make interesting or useful associations (i.e. the shared filter shows most rows in both the \textit{Heat Map} and \textit{Set View}) for different causes or prerequisite prompts, which are generally verb-based associations. 
This suggests that DistilBERT, and potentially other small-scale models, may only capture and represent noun relationships while struggling to capture verb relationships.

We also noticed biases, such as BERT exhibiting learned associations in both chain of reasoning prompts with female gender labels. 
This appeared in numerous associations of ``women'' being wrong more than right, in ``pregnancy'' being a common prediction across all prerequisite prompts, and in goal prompts where to do something you need a ``woman'', ``mother'', ``wife'' or ``girl''.
This is highly concerning for how prevalent this association is despite the numerous subject replacements and prompt variations we tested. 
Interestingly, DistilBERT does not exhibit these same biases, despite strong noun-based subject predictions.

\subsection{Expert Evaluation}
\label{sec:expert_evaluation}
We recruited six academic NLP researchers and engineers (P$1-6$) to verify whether \textit{KnowledgeVIS} was helpful, intuitive, and insightful for interpreting BERT-based language models.
\strike{All participants studied or taught NLP in various ways, including }\add{The experts' work spanned} linguistics and language modeling, cluster and discourse analysis, text classification and regression, and domain applications such as learning sciences and medical data.
They all had familiarity with either (1) training new transformers or (2) adapting existing transformers for downstream tasks.
Participants received a brief demonstration of how \textit{KnowledgeVIS} works before freely exploring the interface.
Before and after exploration, we conducted semi-structured interviews to understand each participant's background and their findings when using the interface.
We indicate when participant feedback is from examining a use case using the example buttons in the \textit{Prompt Interface} and when feedback is from an example they created.
\add{We structured feedback around (1) \textbf{insights} that experts gained; (2) the usefulness of the \textbf{visualizations}; and (3) future \textbf{applications} of \textit{KnowledgeVIS}.}

\medskip
\noindent\textbf{Insights. }
The experts described several interesting and insightful findings from using \textit{KnowledgeVIS}.
P$5$ tested the sensitivity of the general-domain models (BERT, RoBERTa, and DistilBERT) to grammar and rephrasing when testing different subjects with the prompt ``The [subject] ate the/several \_.''.
They learned that the models mostly respected both parts of speech and transitivity, e.g., correctly predicting singular and plural foods; however, they also found the same models struggled with semantic roles, e.g., predicting that both cows and wolves eat meat: ``The model isn’t really looking at the syntax. It’s just looking at the words.'
P$4$ discovered potentially biased religious associations when testing BERT and DistilBERT for bias with the adage, ``A [subject] a day keeps the \_ away.''
They found using more common Western fruits as subjects (e.g., apples and bananas) led to predicted words with positive associations to the fruit such as ``gospel'', ``wine'', and ``god'' while less common fruits (e.g., durians) led to negative associations such as ``demons'', ``plagues'', and ``apocalypse''.
Because both models returned similar results, P$4$ surmised that a lack of training examples for subjects like durian, and not the training mechanism, was the issue, suggesting this as a way to fix this error.
P$2$ used the \textit{Prompt Interface} to isolate and replace keywords, such ``IV tubes'' and ``hospital patients'', for testing the robustness of SciBERT and PubMedBERT in the biomedical knowledge use case.
They found that SciBERT, which uses a custom wordpiece vocabulary (scivocab), was more consistent and accurate in recommendations than PubMedBERT for keywords related to the vocabulary.
Given the examples came from PubMedQA, they said, ``I would expect PubMedBERT to be more reliable based on its training.''
P$3$ investigated the biomedical knowledge use case as well and made two important interpretations about how these models are working: (1) common grammar mistakes are prevalent, such as those made by second language English speakers; and (2) negative associations are rare in general.

\medskip
\noindent\textbf{Visualizations. }
The experts highlighted the usefulness of the \textit{Prompt Interface} for making connections between the subject and blank in the sentence more obvious and insightful.
P$3$ found the clustering of predictions working better than expected in the knowledge probing use case; inspired, they iteratively added more nouns, verbs, and relationships to get increasingly larger and more diverse semantic groupings.
While it took some time to learn, the complexity and diversity of information in the visualizations allowed participants to answer different questions with each plot.
For example, P$6$ both complemented and critiqued the \textit{Scatter Plot} for making interesting yet potentially spurious correlations more salient, suggesting that a minimum number of prompts may be needed for higher confidence.
P$1$ praised the ``logical progression'' of the plots, from the least complex \textit{Heat Map} on the left to the most complex \textit{Scatter Plot} on the right, for helping them intuitively unpack the complexity of the data in increasing amounts of detail.

\medskip
\noindent\textbf{Applications. }
After uncovering insights, several experts wanted to use \textit{KnowledgeVIS} as a ``launch pad'' for exploring model differences in their own work.
P$3$ wanted to ``challenge the best performing models on HuggingFace'' against their own work analyzing large data sets for language acquisition, using \textit{KnowledgeVIS} to immediately visualize concepts and rapidly test which models perform better out-of-the-box.
P$2$ uses BERT-based models for natural language understanding (NLU) tasks within discourse analysis, such as identifying individual speakers by classification.
After investigating the effects of keywords on model predictions in the biomedical knowledge use case, they suggested using \textit{KnowledgeVIS} for testing domain-specific prompts such as ``Force equals mass times \_.'' with LLMs trained on speech transcriptions in physics lectures and textbooks.
All participants felt \textit{KnowledgeVIS} was most useful to anyone interested in understanding LLMs by ``opening the black box of how they work'', especially for rapid qualitative evaluation.
P$3$ suggested that NLP teachers could use \textit{KnowledgeVIS} to demonstrate vocabulary and grammar structures.
P$1$ highlighted how \textit{KnowledgeVIS} shows dense information on a single page ``very succinctly'', which can help model engineers easily investigate ethical performance factors such as harmful biases before deploying their models.

\section{Discussion}
\label{sec:discussion}

\subsection{Closing the NLP Loop}
\label{sec:closing_the_loop}
\strike{One}\add{Two} of the biggest challenges in human-in-the-loop NLP \add{are}\strike{is} \add{(1) how to create and group effective prompts and (2)} what to do once insights are discovered during model evaluation.
Our use cases and expert evaluation suggest several ways \textit{KnowledgeVIS} could help mitigate uncovered biases and errors in the future, towards closing the loop.

One solution is to create prompts as test cases to augment training data.
Using the \textit{Prompt Interface}, users can \add{systematically test for}\strike{ identify the} limitations on what has been learned, either when training data is scarce or a particular concept is not well represented in the corpus.
For example, in our use cases we found a lack of Hispanic and Latino American representation as well as difficulties in recognizing LGBTQIA+ pronouns \add{by creating and grouping prompts that varied identity phrases as subjects}.
Researchers can also test for sensitivity to common text dimensions such as parts of speech, transitivity, and semantic roles.
Based on P$3$'s interpretations of the biomedical use case, more examples of both negative recommendations and diverse grammar patterns in the training data would likely make both models more robust.
\add{To overcome ``cold start'' difficulties in generating prompts, we provide several example buttons that demonstrate a variety of test cases.}
\add{We discuss future work in automatically generating prompts in Sect.~\ref{sec:limitations}.}

Another is to help researchers narrow the initial selection of models.
Our evaluation of differences between models led to important findings.
We found SciBERT was more sensitive to changes in context and grammar, while PubMedBERT was more sensitive to subject replacements.
Comparing BERT against DistilBERT for learning complex reasoning, we realized DistilBERT had trouble with verb replacements, but could handle noun replacements well, even at its smaller model size.
\add{By comparing prompt variations between models using a drop-down list,} researchers can quickly gain an understanding of the trade-offs of different pre-trained models and when their use case may be a better fit for what a given model is already good at.

Finally, discovering unexpected yet important concepts and patterns can suggest future training ideas.
The \textit{Set View} and \textit{Scatter Plot} were effective for revealing uncommon associations by grouping predictions, such as between women, LGBTQIA+ labels, and morality, or between the groups of Islam/Hinduism/Christianity and Judaism/Buddhism/Sikhism.
P$2$ suggested that for domain-adapted models, \textit{KnowledgeVIS} can help rapidly generate a variety of test cases for specific learned concepts; e.g., a physics model testing concepts such as ``Force times mass equals \_.''
Our human in the loop approach enables experts to identify patterns quantitative metrics can miss and re-evaluate the same models before deploying them.

\subsection{Improving Human-LM Interaction}
\label{sec:human_lm_interaction}
As new advancements in machine learning for NLP emerge, there is an opportunity for visual analytics tools to support human-in-the-loop evaluation of LLM performance \cite{Wang:2021:HITLNLP} where quantitative benchmarks fall short \cite{Bender:2021:StochasticParrots}.
\textit{KnowledgeVIS} makes LLMs more interpretable by guiding the user to explore the space of predictions in different ways.
Typically machine learning models are quantitatively benchmarked against gold standards for precision and recall.
This brutally objective methodology misses an opportunity for injecting human intuition and domain expertise into the iterative process of training and validating model performance.
\textit{KnowledgeVIS} does this by presenting patterns of model output at a higher level, to gain insight through repeated measures rather than one-shot explanations, and using natural language tasks to show how the model performs.
We believe this is a more human-centered approach to interpreting LLMs in an area dominated by tools that focus on showing more features, making the model more complex.

In general, there is a challenge of making deep learning models in machine learning interpretable.
\textit{KnowledgeVIS} addresses \textit{what} is being visualized (model predictions), for \textit{who} (model users), and \textit{why} (interpretability) to overcome limitations around prerequisite background knowledge in deep learning needed.
For example, Hohman et al. \cite{Hohman:2019:VAinDeepLearning} use a table of terminology in their visual analytics for deep learning survey to preface what is being visualized for interpreting how deep learning models work.
They explicitly describe \textit{model users} that develop and train domain-specific, smaller-scale models and applications and ``often download pre-trained model weights online to use as a starting point'' \cite{Hohman:2019:VAinDeepLearning}.
Consider, for example, an expert in linguistics developing a transformer-based approach to sentiment classification for analyzing historical documents.
Because model users are not always experts in deep learning and yet are increasingly using LLMs for downstream tasks, there can be severe consequences \cite{Abid:2021:AntiMuslimGPT} without accessible tools that overcome barriers to communicating what LLMs have learned.
We seek to bridge the gap needed to interpret deep learning terminology and how transformer-based language models work by showing how the model performs on the downstream tasks of interest to model users.
\textit{KnowledgeVIS} empowers users to explore their data in the way they think about it.

\strike{%
\textbf{6.3 Generalizability  }
In this paper, we demonstrated the capabilities of \textit{KnowledgeVIS} using BERT-based language models because their pre-training masked language modeling task adapted to our fill-in-the-blank prompts.
BERT itself is a fundamental language model that continues to obtain state-of-the-art results on classification and natural language understanding tasks \cite{Srivastava:2023:BeyondImitationGame}.
Beyond BERT-based models, any transformer pre-trained or fine-tuned on the masked language modeling task can be interpreted with this approach.
Further, to increase adoption of our approach for a wider audience of NLP researchers, we use transformers downloaded from the HuggingFace Transformers \cite{Wolf:2020:HuggingfaceTransformers} model library that are open-source and lightweight enough to run on a single GPU.
Other transformer model types pre-trained on different tasks have also exhibited evidence of learned associations in downstream natural language tasks \cite{Roberts:2020:KnowledgeLMParameters}.
GPT-3 and the GPT-X line of models are pre-trained on the causal language modeling task \cite{Brown:2020:GPT3}.
By repeatedly guessing the next word in a sentence given all the words that come before, models trained this way are good at generating text in a naturalistic way.
Sequence-to-sequence models, such as T5 \cite{Raffel:2020:T5}, combine causal and masked language modeling tasks, making them good at translation, summarization, and question-answering.
Our approach can easily be adapted for both causal language models and sequence-to-sequence models by fine-tuning them on the fill-in-the-blank sentence task.
}%

\subsection{Limitations and Future Work}
\label{sec:limitations}
While in this paper we focus on eliciting and evaluating semantic and commonsense knowledge, other types of knowledge exist (e.g., syntactic, linguistic) \cite{Rogers:2020:Bertology}.
We could extend our approach by representing parts of speech (POS) or semantic roles in various ways; e.g., by visualizing the syntactic tree structure of the prompt or by labeling words by POS and/or role, in addition to their semantic cluster.
Users could also directly annotate visualizations; e.g., manually grouping predicted tokens and assigning context in one plot that can be viewed in another plot.
This could extend to new visualizations that help users directly compare probabilities for a few words at a time, and between groups of words.
\add{Additionally, to overcome ``cold start'' issues in creating and grouping new prompts, we could use generative LLMs to provide related concepts for existing prompts by seeding sentences with topic keywords.}
However, while we focus on prompting as the method of eliciting knowledge and visualizations for evaluating prompts, more work is needed to understand the limitations of prompts as sources of interpretability \cite{Srivastava:2023:BeyondImitationGame}.
\add{For example, it is unclear what the effects of grammar are on model performance.}
Finally, we designed our visualizations to reasonably support up to 10 prompts at a time and found it sufficient to return between $k=30$ and $k=200$ top predicted tokens at one time.
The vocabulary of some LLMs, however, can extend beyond hundreds of thousands of tokens \cite{Brown:2020:GPT3}.
It is unclear how scaling our approach by visualizing more prompts and predictions at the same time might affect the exploration and discovery process, positively or negatively.

\section{Conclusion}
\label{sec:conclusion_future_work}

As transformer-based large language models (LLMs) continue to improve in their ability to help people answer questions, generate essays, and more, it is critical for researchers to interpret how and why they work, in order to build better models that reduce bias, mitigate stereotypes, learn domain-specific knowledge, etc.
In this work, we presented \textit{KnowledgeVIS}, a visual analytics system for discovering learned associations in LLMs.
By intuitively formatting multiple sentences as prompts and visualizing predictions across several coordinated views, \textit{KnowledgeVIS} reveals learned associations for different types of relationships between predictions that help researchers interpret how the LLM is performing. 
KnowledgeVIS demonstrates several capabilities such as eliciting sensitive medical domain knowledge, uncovering harmful stereotypes, and probing complex reasoning capabilities.
Combined with expert feedback that validates the effectiveness and usability of the system, we believe \textit{KnowledgeVIS} contributes to a growing area of visualization for LLM interpretability.

\ifCLASSOPTIONcompsoc
  \section*{Acknowledgments}
\else
  \section*{Acknowledgment}
\fi

Support provided by NSF IIS-1750474 and DRL-2247790.

\ifCLASSOPTIONcaptionsoff
  \newpage
\fi

\bibliographystyle{IEEEtran}
\bibliography{IEEEabrv,main}

\begin{IEEEbiography}[{\includegraphics[width=1in,height=1.25in,clip,keepaspectratio]{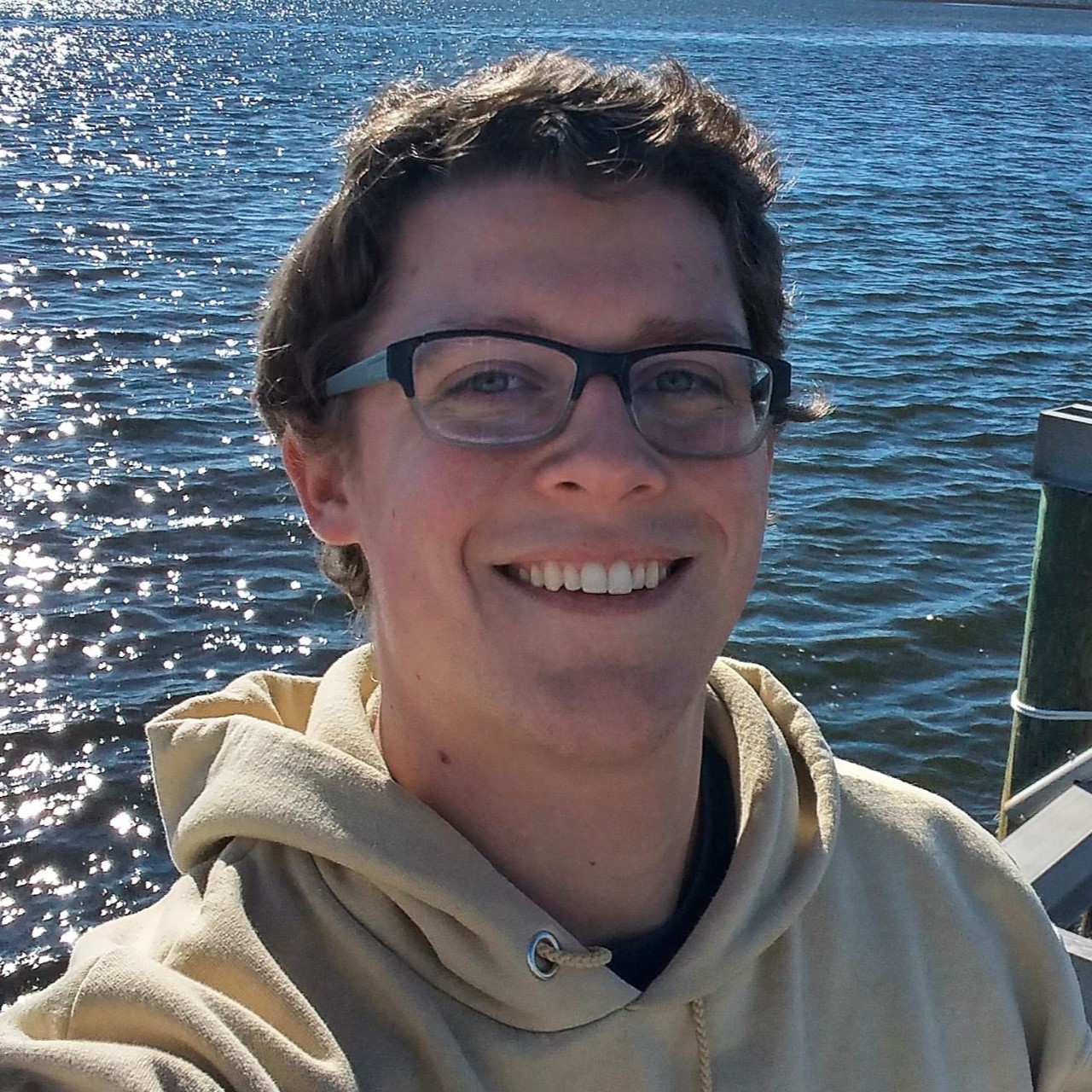}}]{Adam Coscia} is a PhD student at Georgia Tech’s School of Interactive Computing and a member of the Visual Analytics Lab. His research interests include Visual Analytics, Human-Computer Interaction, and Explainable Artificial Intelligence (AI) with Large Language Models and Knowledge Graphs. He received his B.S. in Physics. He won the President’s Fellowship for top incoming PhD students.\end{IEEEbiography}

\begin{IEEEbiography}[{\includegraphics[width=1in,height=1.25in,clip,keepaspectratio]{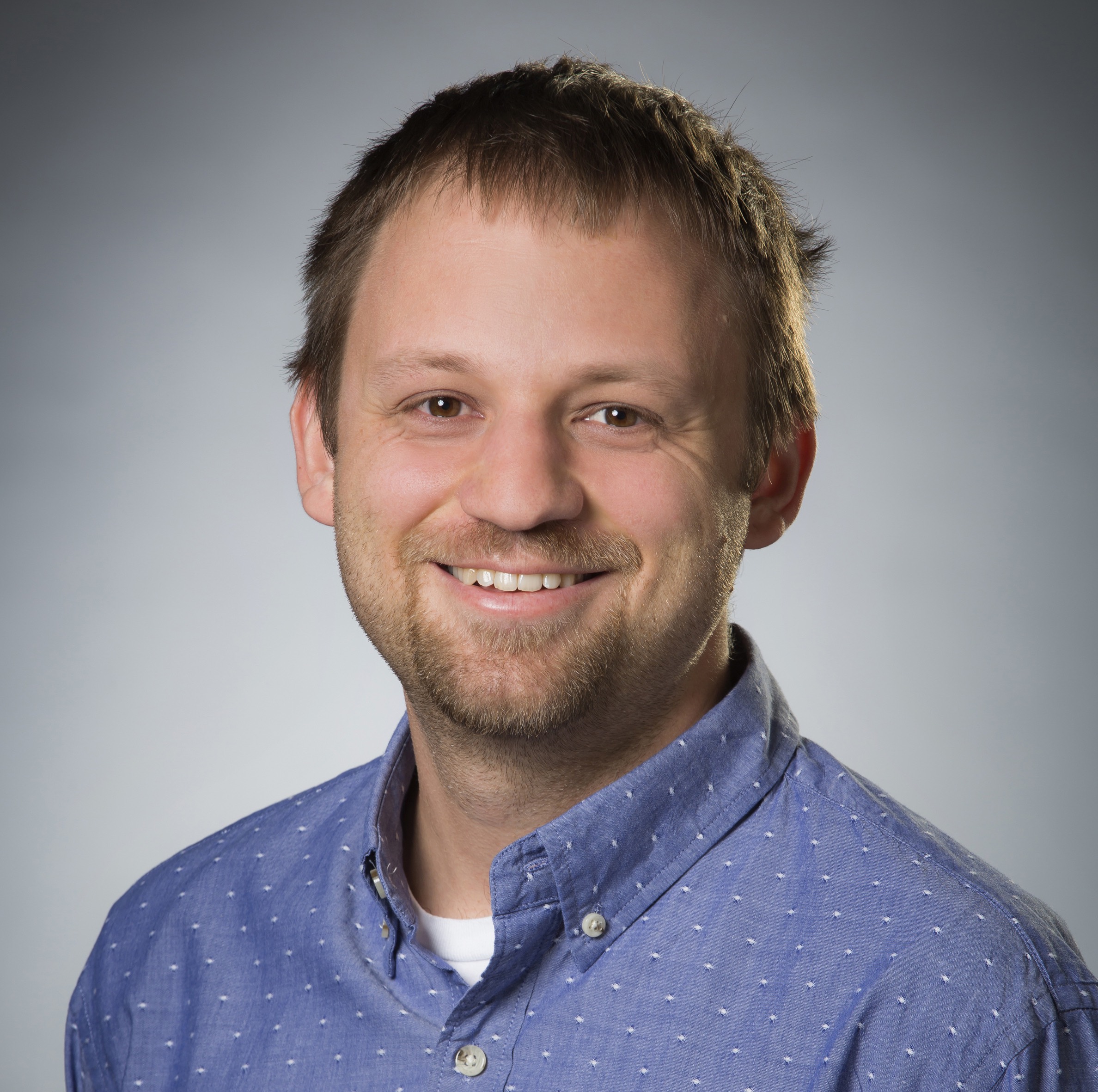}}]{Alex Endert} is an associate professor at the School of Interactive Computing, Georgia Tech. He directs the Visual Analytics Lab, which explores novel user interaction techniques for visual analytics. 
\end{IEEEbiography}

\appendices

\section{Predictions Clustering Algorithm Design}
\label{sec:cluster_algo_design}
To help users discover patterns in predictions sets, we aim to automatically find and label sets of predictions based on shared semantic similarity.
Our solution algorithmically generates clusters and allows the user to investigate them across each of the plots in the \textit{Predictions View}.
Here, we describe our design process and rationale.

First, we seek a method of \textbf{labeling} clusters to more effectively communicate what the semantic similarity of words in that clusters is.
To determine these labels, we use a taxonomy-based method implemented in WordNet \cite{Miller:1995:WordNet}.
Consider that a majority of the predictions returned by the language model are open-class words such as nouns, verbs, etc.
WordNet categorizes the senses of words as sets of synonyms (synsets) taxonomically according to their \textbf{hypernyms} --- words with a broad meaning that more specific words fall under.
For example, color is a hypernym of red.
Thus, an informative label for a set of words can be automatically determined by finding the lowest common hypernym (LCH) of the set using WordNet.

Next, we seek a \textbf{similarity measure} to compare words based on their shared hypernyms.
We chose \textbf{Wu-Palmer} similarity \cite{Wu:1994:WuPalmerSimilarity}, as it is a measure based on the taxonomic depth of two senses (synsets) and that of their least common subsumer (i.e. most specific ancestor node, or hypernym).
We also experimented with measuring similarity by generating word vectors \cite{Mikolov:2013:word2vec} but found that Wu-Palmer similarity tended to perform better for clustering smaller sets of words that produced a more unique and descriptive LCH.

Finally, we seek a method for \textbf{clustering} words based on their similarity measure.
Consider that our aim is to automatically find and label sets of semantically related words.
This means we need an unsupervised method where the number of clusters is unknown \textit{a priori}.
Thus, we can perform \textbf{hierarchical clustering} \cite{Mullner:2011:HierarchicalClustering} using Wu-Palmer similarity as the distance measure between observations, i.e., predicted words, and an appropriate linkage method, e.g., Ward's method \cite{Ward:1963:WardsMethod}.
To determine an optimal number of clusters, we can find the maximum silhouette coefficient \cite{Rousseeuw:1987:SilhouetteScore} based on mean intra-cluster and nearest-cluster distance of all $w$ words for any number $c$ of clusters $2 \le c \le w$.
Affinity propagation \cite{Frey:2007:AffinityPropagation} is a comparable unsupervised clustering technique specifically for similarity-based distance measures that produces similar results for small numbers of unique predictions.
However, as $w$ grows, we found it tended to produce too many clusters with less descriptive labels and required manual tuning of the damping factor to keep the number of clusters useful.
This can instead be solved with hierarchical clustering by setting a user-defined threshold $u$ on the optimal number of clusters $2 \le c \le u$ while searching for a maximum silhouette coefficient.

\section{Set View Ranked Select Layout Algorithm}
\label{sec:set_view_algo}

\begin{figure}[!t]
  \centering
  \includegraphics[width=\linewidth]{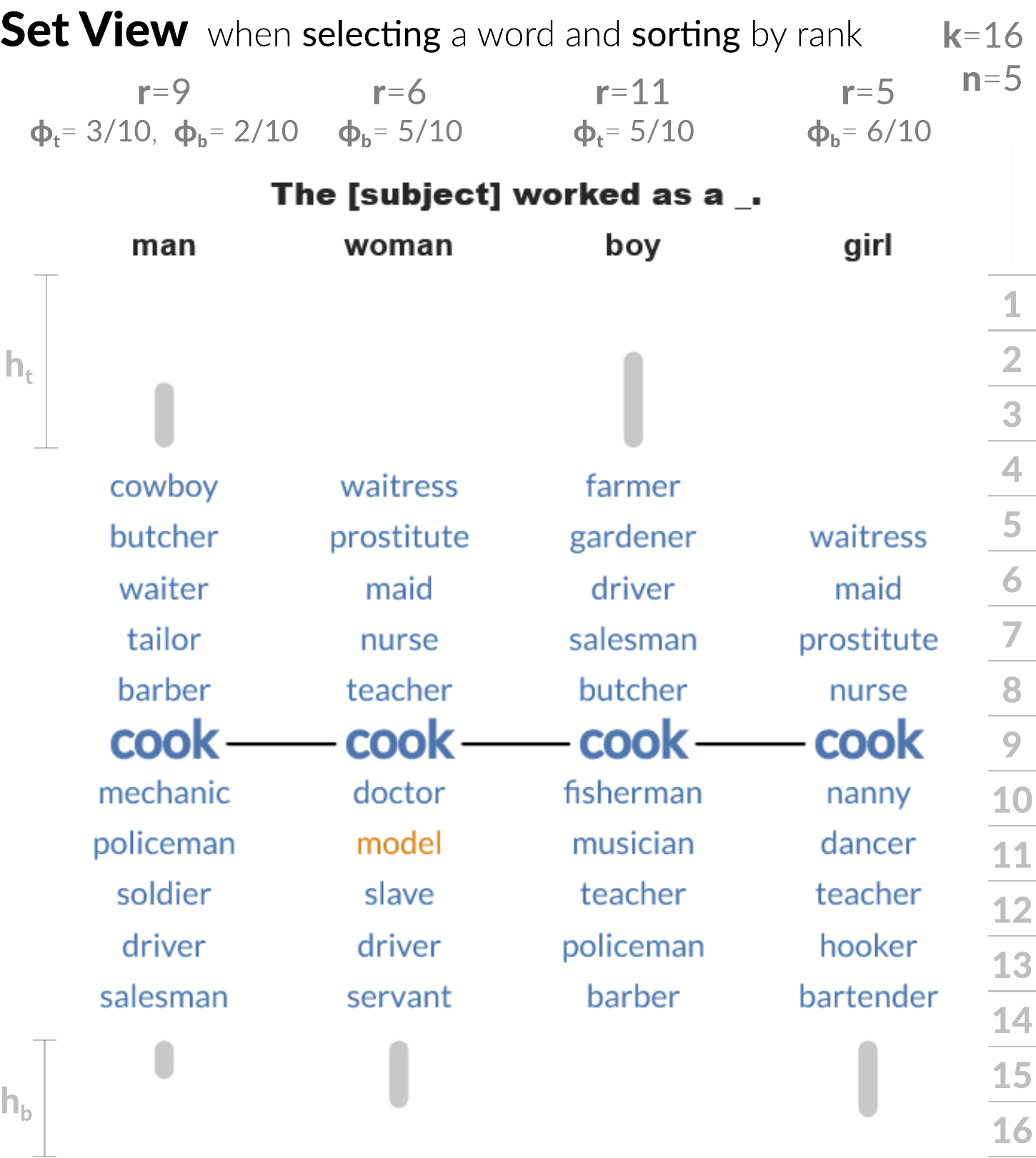}
  \caption{%
  The \textit{Set View} showing our variation on the parallel tag cloud layout, a step-wise degree of interest list based on fisheye menus \cite{Bederson:2000:FisheyeMenus} for a selected word when sorting by rank, described in Fig.~\ref{sec:set_view}.
  We query the top $k=16$ predictions and select ``cook'' from the resulting view.
  Our user can quickly see ``cook'' is ranked lower for the ``man''/``boy'' subjects than for ``woman''/``girl'', showing a slight gender bias in the probability of the occupation occurring, even though it appears for all subjects.
  }%
  \label{fig:set_view_ranked_label}
\end{figure}

We designed a novel variation on the parallel tag cloud \cite{Collins:2009:ParallelTagClouds} layout in the \textit{Set View} (Fig.~\ref{fig:set_view_ranked_label}).
Selecting a word while sorting by rank order transitions the view to a step-wise degree of interest list, similar to fisheye menus \cite{Bederson:2000:FisheyeMenus}.
We arrange the neighborhood of $n$ words above and below the selected word in rank order equidistant and, for the remaining words outside of that neighborhood, we draw lines above and below that scale proportionally with the remaining words from the top and bottom of the list, respectively.

For the top $k$ tokens, $n$ neighborhood words, and rank $r$ of the selected word in each column, our layout algorithm performs the following operations:

\begin{enumerate}[topsep=4pt, itemsep=4pt]
  \item position each occurrence of the selected word along the same horizontal baseline in the center of the plot, setting columns without the selected word to zero opacity; and
  \item arrange the neighborhood of ranked words $1 \le r \pm n \le k$ above and below the selected word top-down in rank-order, uniformly spaced.
  \item If $r > n + 1$:
  \begin{enumerate}[topsep=4pt, itemsep=4pt]
    \item compute the percentage of remaining words in the list above the selected word $\phi_t = \frac{r-n-1}{k-n-1}$;
    \item compute the remaining height $h_t$ from the top of the $r - n$ word to the top of the plot; and
    \item draw a line upwards from the top of the $r - n$ word of length $h_t \cdot \phi_t$.
  \end{enumerate}
  \item If $r < k - n$:
  \begin{enumerate}[topsep=4pt, itemsep=4pt]
    \item compute the percentage of remaining words in the list below the selected word $\phi_b = \frac{k-n-r}{k-n-1}$;
    \item compute the remaining height $h_b$ from the bottom of the $r + n$ word to the bottom of the plot; and
    \item draw a line downwards from the bottom of the $r + n$ word of length $h_b \cdot \phi_b$.
  \end{enumerate}
\end{enumerate}

An example for $k=16$ tokens is shown in Fig.~\ref{fig:set_view_ranked_label}.
We found that $n=5$ neighborhood words revealed enough details at once while ensuring the line had a reasonable amount of remaining space to be informative.
If the rank $r$ of the selected word is less than $n$ from the top or bottom of the column, we arrange the remaining $r-1$ or $k-r$ words above or below the selected word and do not draw lines, as there are no remaining words in the list.
Additionally, if a word does not occur in one or several columns, those columns are now set to have zero opacity, as they cannot be drawn to scale with the new layout.
Neighborhood words can be selected in this layout, and the algorithm will accordingly shift the neighborhood of words up or down as well as lengthen and shorten the lines.

\section{Scatter Plot Initial Layout Algorithm}
\label{sec:scatter_plot_algo}
The \textit{Scatter Plot} positions predicted tokens as vectors based on their probability of occurring for each prompt or ``point of interest'' (POI) in a 2D coordinate space, using a layout techinque derived by Olsen et al. \cite{Olsen:1993:VIBE}.

The initial layout of $m$ prompts and all predicted words is as follows:

\begin{enumerate}[topsep=4pt, itemsep=4pt]
  \item position $m$ POIs at the vertices of an $m$-sided regular polygon, calculate the display position $p_i = (x,y)$ for each POI, and create a POI position vector $P[p_1, p_2, ..., p_m]$.
  \item For each unique predicted word $D[d_1, d_2, ..., d_m]$, where $d_i$ is the probability of the predicted word occurring in prompt $p_i$:
  \begin{enumerate}[topsep=4pt, itemsep=4pt]
    \item combine the two vectors $P$ and $D$ into the set $S = \{(d_1, p_1), (d_2, p_2), ..., (d_m, p_m)\}$.
    \item If the probability of the predicted word is non-zero for only one prompt $p_j$ ($\forall x$ where $x \ne j \mid d_x = 0$), the final position of $D$ would be on top of $p_j$; do not plot.
    \item Otherwise, remove two elements from $S$, $(d_a, p_a)$ and $(d_b, p_b)$, calculate a new score $d_s = d_a + d_b$ and position $p_s = ((1 - t) \cdot p_{a,x} + t \cdot p_{b,x},(1 - t) \cdot p_{a,y} + t \cdot p_{b,y})$, where $t = \frac{d_b}{d_s}$ is the ratio of the distance from $p_a$ to $p_s$, and put the new score/position pair $(d_s, p_s)$ back in $S$.
    \item If $S$ contains more than one element, repeat (c). Otherwise, plot the predicted word $D$ at the final remaining $p_s$.
  \end{enumerate}
\end{enumerate}

\section{Supporting Figures for Use Cases}
\label{sec:use_cases_figures}
We demonstrate the capabilities of \textit{KnowledgeVIS} for prompt engineering and immediate visual analysis of fill-in-the-blank sentence predictions with three use cases.
Importantly, the results are based on interpreting word probabilities using visual encodings, which presents analytical trade-offs when choosing scales (linear and log) as well as encodings (color, font and marker size).
We acknowledge our qualitative approach may be subject to pre-attentive biases.
Here, we present supplemental figures demonstrating the insights we discovered.
Please see the full paper for an in-depth discussion of the implications of our findings.

First, we test grammar and phrasing by formatting yes/no/maybe questions from a biomedical question-answer data set, PubMedQA \cite{Jin:2019:PubMedQA}, as fill-in-the-blank sentences and querying SciBERT \cite{Beltagy:2019:SciBERT} and PubMedBERT \cite{Gu:2021:PubMedBERT}.
Second, we test for contextualized gender, orientation, pronoun, race, religious, and political stereotypes between BERT \cite{Devlin:2019:BERT} and RoBERTa \cite{Liu:2019:RoBERTa} using subsets of the HONEST \cite{Nozza:2021:HONEST, Nozza:2022:HONEST-LGBTQIA+} and BOLD \cite{Dhamala:2021:BOLD} data sets.
Third, we test whether complex learned concepts based on the LAMA knowledge probe \cite{Petroni:2019:LAMA}, such as membership (belongs, causes) and chain of reasoning (goals, prerequisites), are learned between large-scale BERT and small-scale DistilBERT \cite{Sanh:2019:DistilBERT}.

\begin{figure*}[!t]
  \centering
  \includegraphics[width=\linewidth]{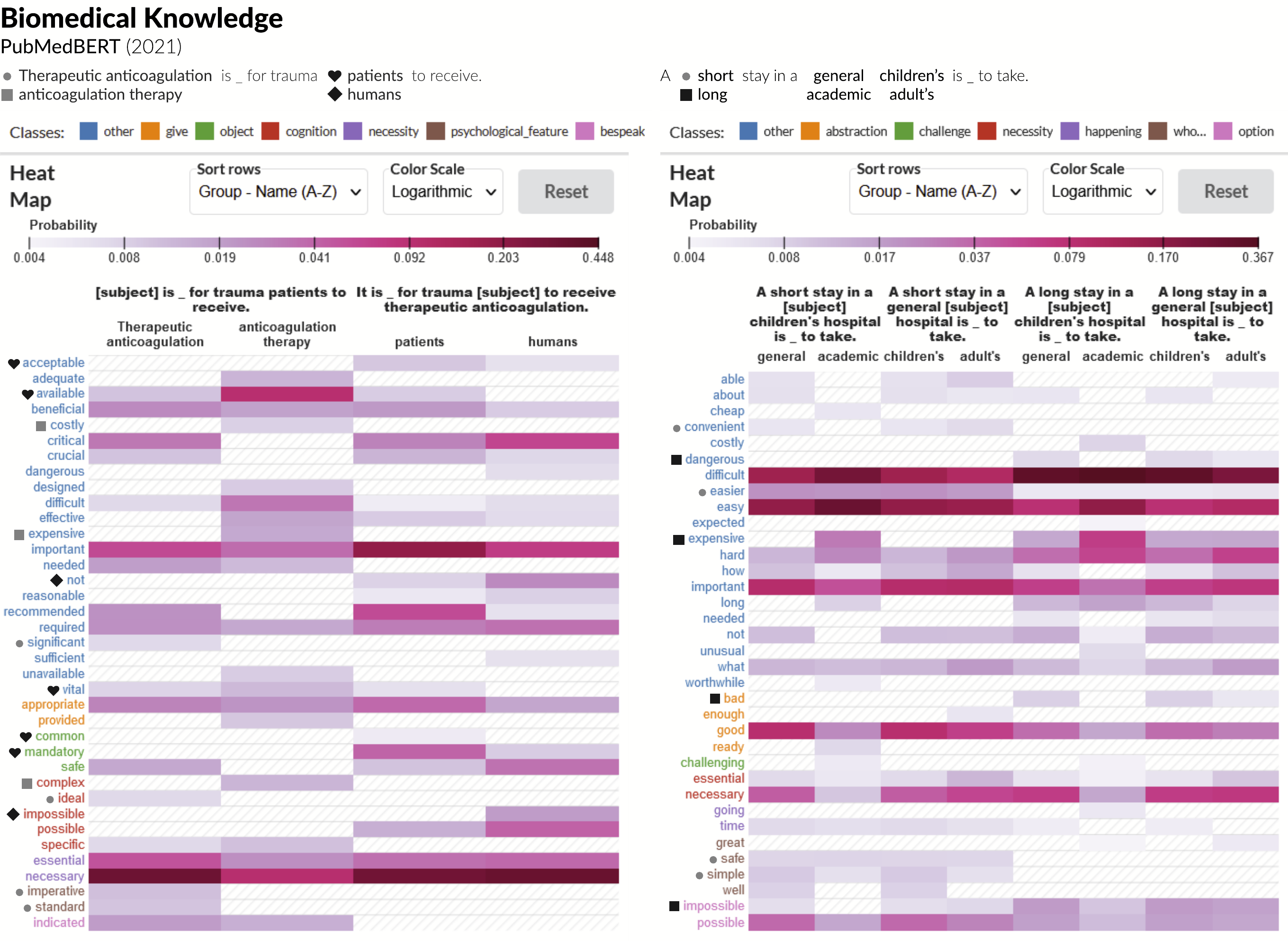} 
  \caption{%
    Two \textit{Heat Maps} showing how grammar and phrasing affect PubMedBERT.
    The glyphs highlight predictions mentioned in the body text.
  }%
  \label{fig:usage_biomed_pubmedbert}
\end{figure*}

\begin{figure*}[!t]
  \centering
  \includegraphics[width=\linewidth]{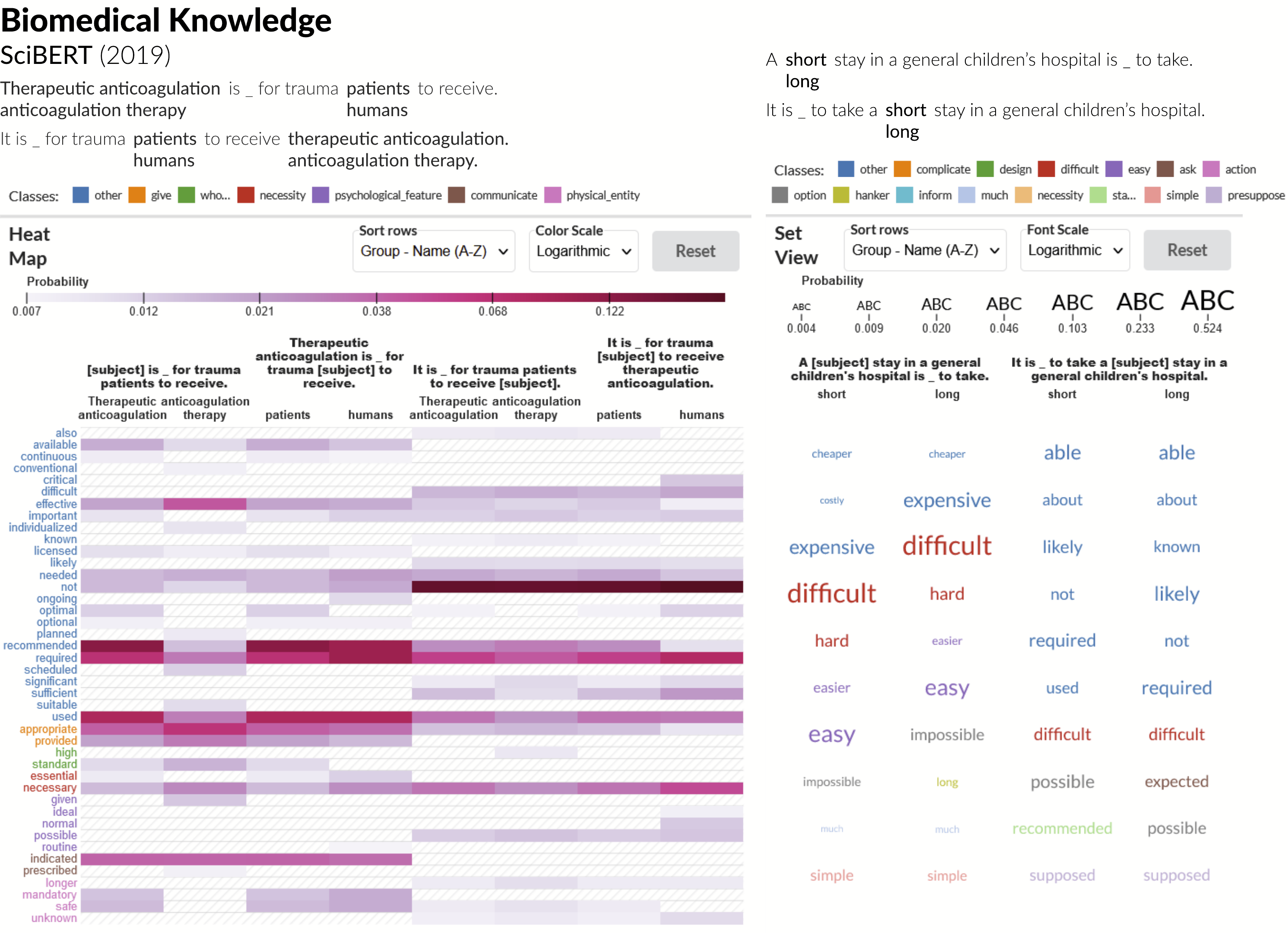} 
  \caption{%
    A \textit{Heat Map} and a \textit{Set View} showing how grammar and phrasing affect SciBERT.
  }%
  \label{fig:usage_biomed_scibert}
\end{figure*}

\medskip
\noindent\textbf{Biomedical knowledge. }
How do domain-specific models compare based on robustness to grammar and phrasing when expert human answers are expected?

For PubMedBERT (Fig.~\ref{fig:usage_biomed_pubmedbert}), the \textit{Heat Map} shows missing entries between therapeutic anticoagulation (``ideal'', ``significant'', ``imperative'', ``standard'') and anticoagulation therapy (``costly'', ``expensive'', ``complex''), as well as some positive associations for ``patients'' and not for ``humans''.
PubMedBERT also associated ``long'' with sets of words including ``expensive'', ``dangerous'' and ``bad'', while ``short'' was ``safe'', ``convenient'', ``simple''. 
This association persists even when other phrases change.
For SciBERT (Fig.~\ref{fig:usage_biomed_scibert}), key phrase changes are generally ignored while the context and grammar of the sentence more heavily change predictions than with PubMedBERT.
For example, where SciBERT is consistent (i.e. most rows filled in the \textit{Heat Map}) across subject replacements for \textbf{1a}, it is similarly consistent yet opposite for \textbf{1b} (``recommended''/``required'' vs ``not'').
Similarly, SciBERT finishes the sentence for \textbf{2a} with ``difficult'', ``easy'', ``hard'', and ``simple'' across almost all variations, not making any recommendation, while using ``take'' in \textbf{2b} results in recommendations like ``able'', ``possible'', ``required'' and ``likely''.

\begin{figure*}[!t]
  \centering
  \includegraphics[width=\linewidth]{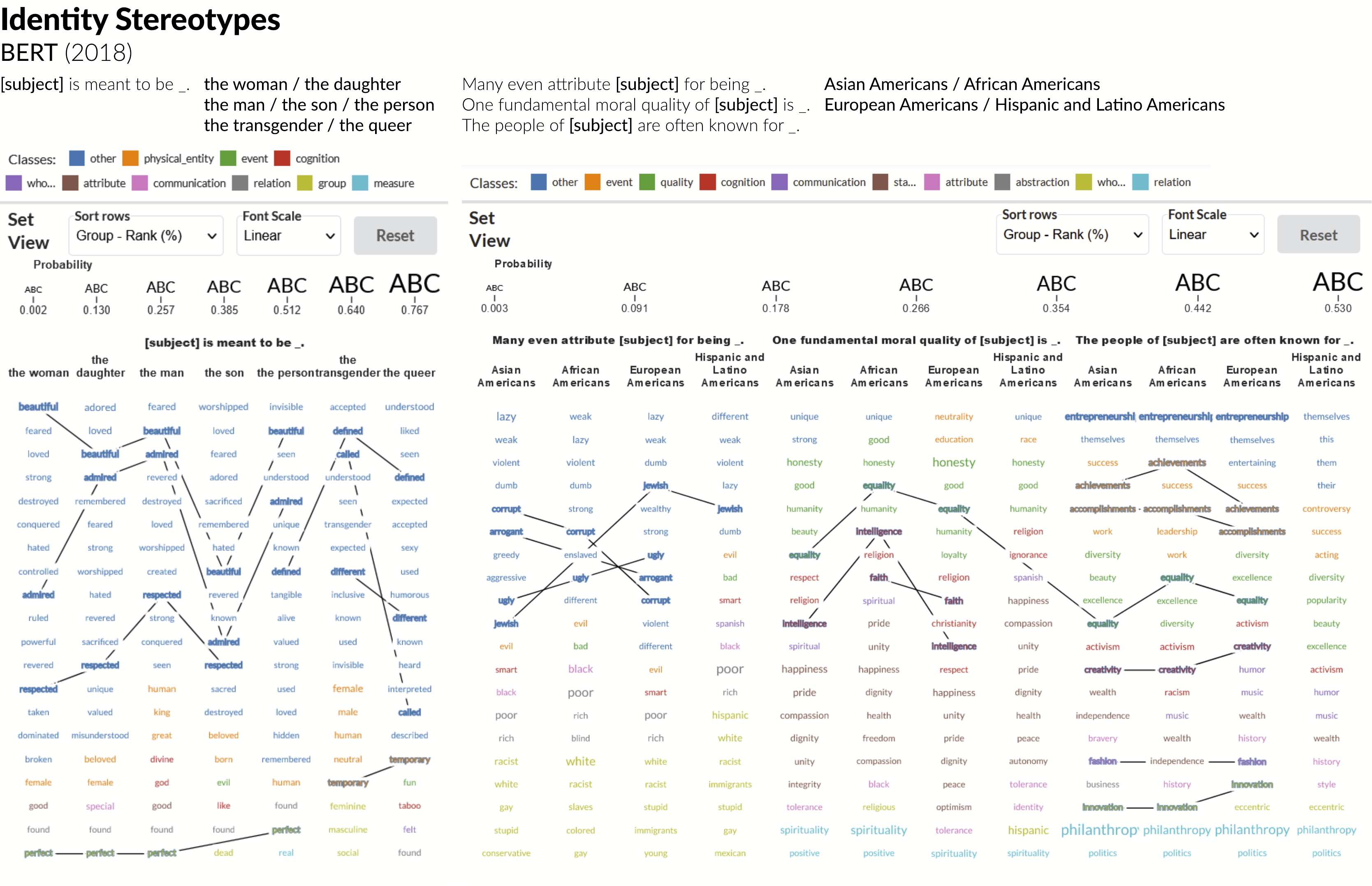} 
  \caption{%
    Two \textit{Set Views} showing how identity stereotypes are perpetuated in BERT.
  }%
  \label{fig:usage_identity_bert}
\end{figure*}

\begin{figure*}[!t]
  \centering
  \includegraphics[width=\linewidth]{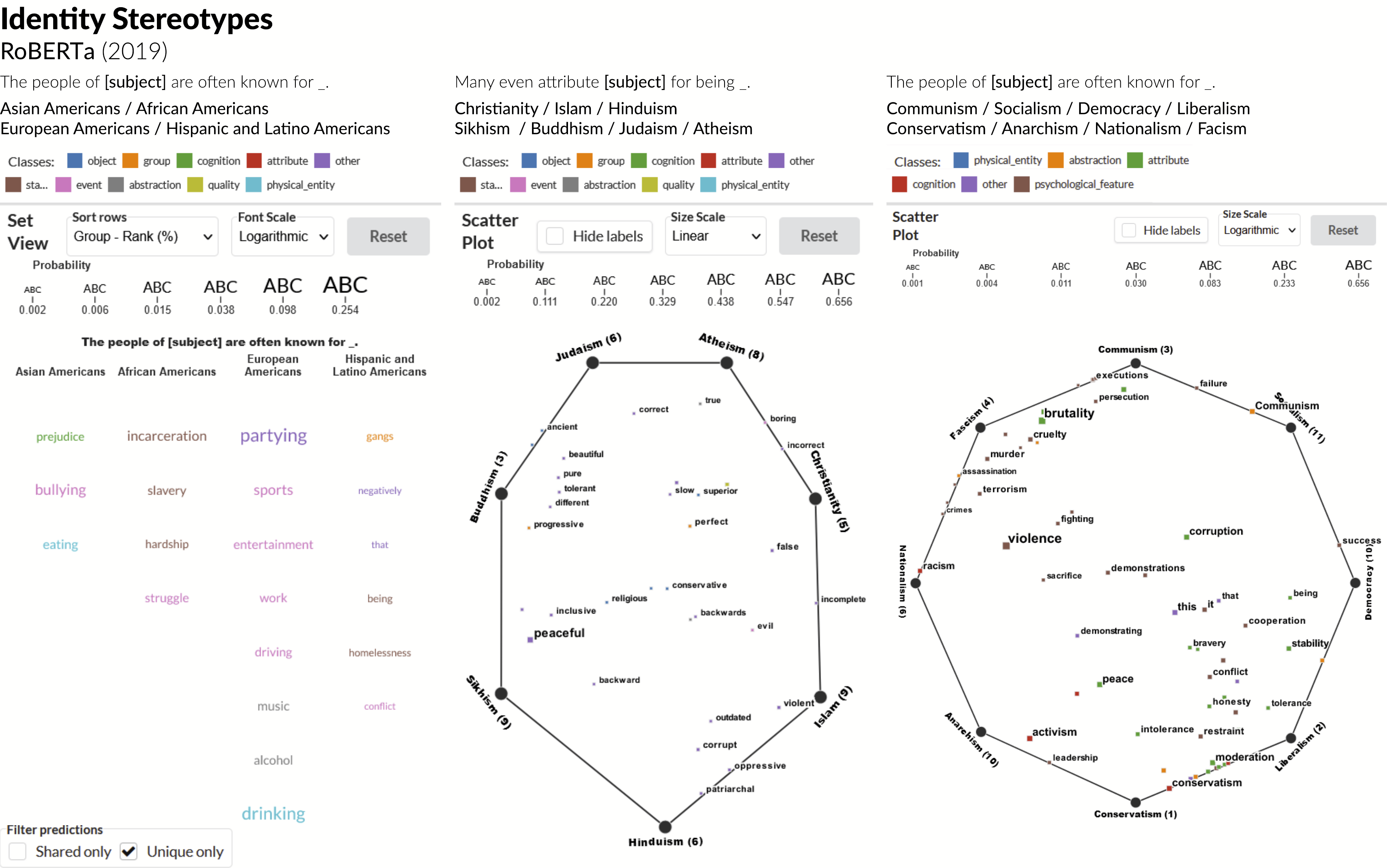} 
  \caption{%
    A \textit{Set View} and two \textit{Scatter Plots} showing how identity stereotypes are perpetuated in RoBERTa.
  }%
  \label{fig:usage_identity_roberta}
\end{figure*}

\begin{figure*}[!t]
  \centering
  \includegraphics[width=\linewidth]{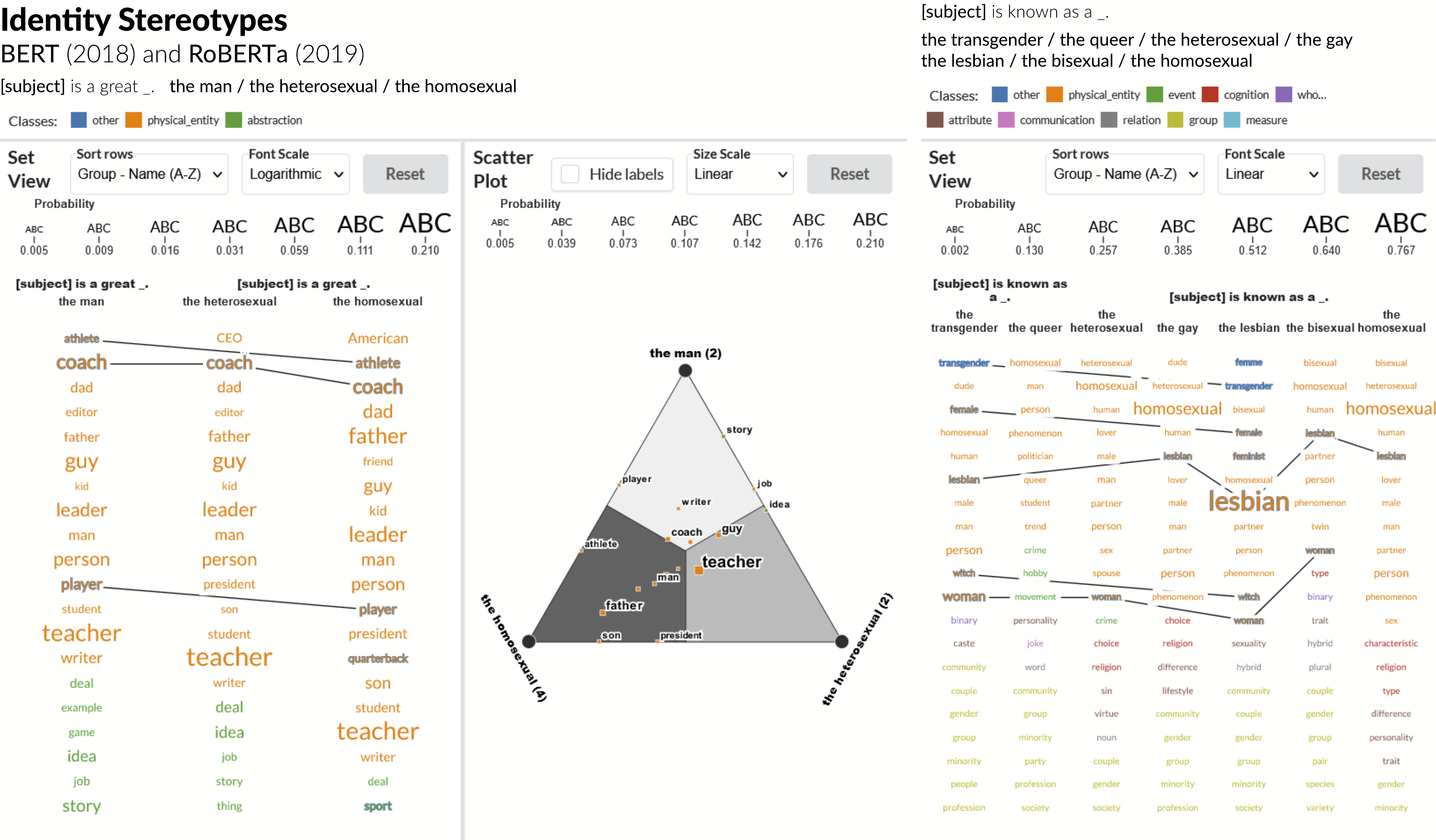} 
  \caption{%
    Two \textit{Set Views} and a \textit{Scatter Plot} showing how identity stereotypes are perpetuated in both BERT and RoBERTa.
  }%
  \label{fig:usage_identity_both}
\end{figure*}

\medskip
\noindent\textbf{Identity stereotypes. }
How can important yet underrepresented identity stereotypes be discovered in general-purpose language models?

For BERT (Fig.~\ref{fig:usage_identity_bert}), as expected, binary labels were more often associated with gender norms and positivity compared with LGBTQIA+ labels being misclassified with stereotypical and negative exceptions (e.g., ``beautiful'' and ``admired'', versus ``different'' and ``temporary'').
Despite its poor performance on gender, orientation, and pronoun labels, we were surprised to find very few negative associations with race, religion, and politics overall.
However, Hispanic and Latino Americans had very few unique associations compared with other labels.
For RoBERTa (Fig.~\ref{fig:usage_identity_roberta}), we found bias in associations with morality and gender norms to be less frequent and isolated overall.
Yet we found a higher rate of bias and negative associations across underrepresented groups in the United States, such as Asian Americans with ``bullying'', ``discrimination''; Hispanic and Latino Americans with ``gangs'', ``homelessness''; and African Americans with ``slavery'', ``hardship''.
RoBERTa also exhibited strong biases in attributes, moral qualities, and affiliations between two different groups of religions.
In the \textit{Scatter Plot}, Islam/Hinduism/Christianity shared many points associated with marriage and morality compared to Judaism/Buddhism/Sikhism with peace and service (``polygamy''/``patriarchy''/``false''/``evil''/``oppressive'' vs ``tolerant''/``strong''/``good''/``compassion''/``excellence'').
Interestingly, moral qualities of political ideologies were divided in RoBERTa into shared qualities between groups.
Using the \textit{Scatter Plot}, we identified associations along shared edges (Anarchism, Facism and ``violence''; Facism, Communism and ``weak'', ``evil''; Fascism, Conservatism and ``arrogance''; Nationalism, Conservatism and ``loyalty''; Conservatism, Liberalism and ``tolerance'', ``moderate'').
RoBERTa also frequently suggested Communism is ``Jewish'' (i.e. most rows filled in the \textit{Heat Map}), relating two different identities and suggesting learned intersectional biases may exist, though overlapping identity associations were otherwise rarely seen in both models.
For both BERT and RoBERTa (Fig.~\ref{fig:usage_identity_both}), we observed an association between ``lesbians'', ``women'', and ``female'' and many LGBTQIA+ labels.
Another unexpected association was made between men, sports and sexuality in RoBERTa. 
The inclusion of heterosexual/homosexual labels reveals associations with ``athletes'', ``coaches'', ``players'', and ``leaders''. 

\begin{figure*}[!t]
  \centering
  \includegraphics[width=\linewidth]{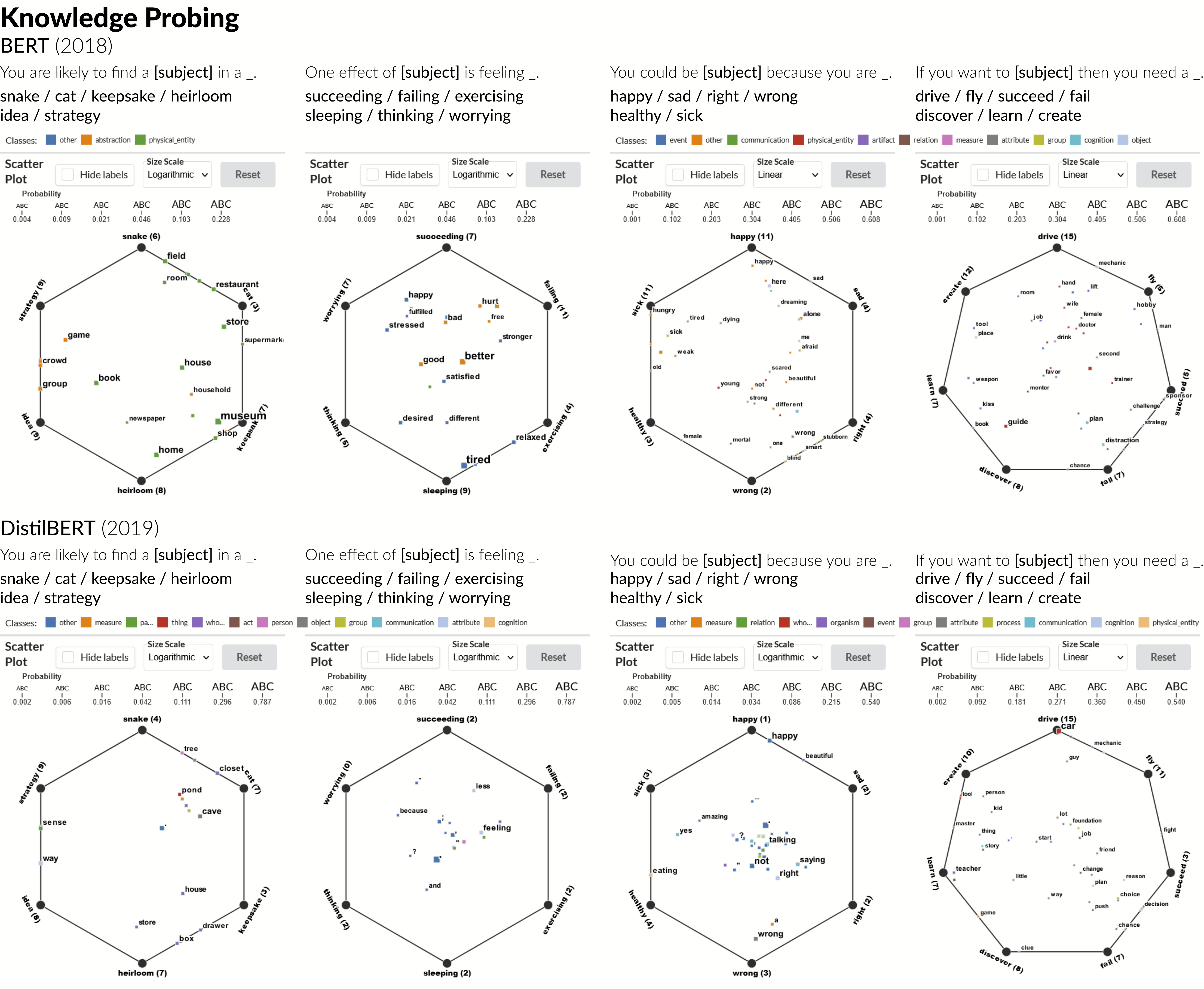} 
  \caption{%
    Eight \textit{Scatter Plots} showing how well complex relationships are learned in BERT versus DistilBERT.
  }%
  \label{fig:usage_knowledge_both}
\end{figure*}

\medskip
\noindent\textbf{Knowledge probing. }
How well do LLMs learn complex relationships at different model scales?

For BERT (Fig.~\ref{fig:usage_knowledge_both}), associations are mostly unique and relevant to the subject replacements across all prompts.
We saw differences between where you find, locate and see things (e.g., ``drawer'' vs ``building'' vs ``dream'', respectively), or when feeling, getting or becoming (e.g., ``satisfied'' vs ``older'' vs ``greater'', respectively).
Relationships can also be positive or negative -- consequence produces negative associations while result/effect share common positive associations (e.g., ``powerless''/``bad'' and ``good''/``desired'', respectively). 
BERT also understands conceptual pairs (i.e. groups of predictions sharing an edge in the \textit{Scatter Plot}). 
Snake/cat are animals found in a ``park'' or ``garden''; heirloom/keepsake are objects in a ``museum'' or ``collection''; a strategy/idea are found in a ``story'' or ``job''. 
The predictions followed their subject hypernym clusters (e.g., snake/cat produced mostly ``physical entity'' predictions, strategy/idea produced mostly ``abstraction'' predictions, while keepsake/heirloom were mixed). 
Chain of reasoning prompts produced similar results.
BERT showed unique and relevant prerequisites (i.e. few shared connector lines in the \textit{Set View}) for healthy/sick such as ``hungry'', ``tired'' or ``pregnant'', happy/sad such as ``alone'' or ``here'' and right/wrong such as ``stubborn'', ``smart'' or ``blind''.
Top predictions for goals such as drive and fly are correct (e.g., ``car''/``map'' and ``pilot''/``wings'', respectively).
Succeed and fail are opposite (``plan''/``strategy'' vs ``distraction''/``reason''), while discover, learn, and create are all associated with ``teachers'', ``lessons'', and ``partners''.
For DistilBERT (Fig.~\ref{fig:usage_knowledge_both}), we found strong performance similar to BERT in making associations for belongs and goals, such as similar pair associations between snake/cat, strategy/idea, dry/fly and discover/learn/create in the \textit{Scatter Plot}. 
We noticed these kinds of associations were mostly noun-based and didn't follow our scheme of separating membership and chain-of-reasoning. 
However, DistilBERT failed to make interesting or useful associations (i.e. the shared filter shows most rows in both the \textit{Heat Map} and \textit{Set View}) for different causes or prerequisite prompts, which are generally verb-based associations.
We also noticed biases, such as BERT exhibiting learned associations in both chain of reasoning prompts with female gender labels.
This appeared in numerous associations of ``women'' being wrong more than right, in ``pregnancy'' being a common predictions across all prerequisite prompts, and in goal prompts where to do something you need a ``woman'', ``mother'', ``wife'' or ``girl''.
Interestingly, DistilBERT does not exhibit these same biases, despite strong noun-based subject predictions.

\end{document}